\documentclass[10pt]{article}
\usepackage[latin1]{inputenc}
\usepackage{amsmath}
\usepackage{amsfonts}
\usepackage{amssymb}
\usepackage{subcaption}
\usepackage{mathrsfs}
\usepackage{afterpage}
\usepackage[table,xcdraw]{xcolor}
\usepackage{tabularx}
\usepackage{comment}
\usepackage{bm}
\usepackage{cancel}
\usepackage{ulem}
\usepackage{blindtext}
\usepackage{pdfpages}
\usepackage{MnSymbol}
\usepackage{hyperref} 
\usepackage[cal=boondox,scr=boondoxo]{mathalfa}
\usepackage{float}
\usepackage{fancybox,graphicx}
\usepackage{caption}
\usepackage{color}
\usepackage{authblk}
\usepackage{cancel} 
\usepackage{accents}
\usepackage[titletoc,title]{appendix}
\usepackage{cite}
\usepackage{multirow}
\usepackage{makecell}
\usepackage{mathtools}
\usepackage{sidecap}
\usepackage{footmisc}

\newcommand{\bqa}{\begin{eqnarray}}
\newcommand{\eqa}{\end{eqnarray}}
\newcommand{\beq}{\begin{equation}}
\newcommand{\eeq}{\end{equation}}

\usepackage[top=2in, bottom=1.5in, left=1in, right=1in]{geometry}

\usepackage{stackengine}

\def\lQ{\Lambda_{\rm QCD}}
\def\als{\alpha_s} 
     
\def\j{{\cal J}}

\def\m{{\cal M}}
\newcommand{\be}{\begin{equation}}
\newcommand{\ee}{\end{equation}}
\newcommand{\bea}{\begin{eqnarray}}
\newcommand{\eea}{\end{eqnarray}}
\newcommand{\nn}{\nonumber}

\title{Hybrid to Quarkonia transitions} 
\author[1]{Rub\'en Oncala}
\author[2,3]{Joan Soto}
\affil[1]{Department of Mathematics and Data Science, Universidad San Pablo-CEU, \newline CEU Universities, Calle Juli\'an Romea 23, 28003 Madrid, Spain.}
\affil[2]{Departament de F\'isica Qu\`antica i Astrof\'isica and Institut de Ci\`encies del Cosmos, Universitat de Barcelona, Mart\'i i Franqu\`es 1, 08028 Barcelona, Catalonia, Spain.}
\affil[3]{Institut d'Estudis Espacials de Catalunya (IEEC), 08860 Castelldefels, Catalonia, Spain.}

\begin{document}
    \maketitle
\begin{abstract}
Hybrid quarkonia -exotic hadrons with explicit gluonic degrees of freedom- have gained increasing attention in hadron spectroscopy, particularly with the ongoing discovery of new XYZ mesons. In this work, we update the spectrum of heavy hybrid mesons in the charmonium and bottomonium sectors using the Born-Oppenheimer Effective Field Theory framework, by incorporating the latest lattice QCD results for hybrid static potentials.
We refine earlier calculations and analyze allowed transitions from hybrids to conventional quarkonia, including both spin-conserved and spin-flip decays. We carry out a comprehensive error analysis and discuss the reliability of our results. We compare them to experimental data of the Particle Data Group, which allows us to identify hybrid candidates among the observed XYZ states. We provide hybrid or quarkonium interpretations for nearly all heavy isospin-zero mesons observed and incorporate new hybrid candidates.
\end{abstract}

\clearpage
    
\section{Introduction}
\label{sec:intro}

Understanding the nature of exotic hadrons is one of the key challenges in modern hadron spectroscopy. The quark model originally formulated by Gell-Mann \cite{Gell-Mann:1964ewy} and Zweig \cite{Zweig:1964ruk} classifies hadrons into mesons (quark-antiquark states) and baryons (three-quark states). However, Quantum Chromodynamics (QCD) permits more complex configurations, such as hybrid mesons that incorporate explicit gluonic degrees of freedom \cite{Jaffe:1975fd,Morningstar:1999rf}. The presence of an excited gluon field endows hybrids with quantum numbers that are distinct from those of conventional quarkonia, making their experimental identification and theoretical description especially compelling.\\

In recent years, several exotic candidates collectively known as the XYZ states have been observed \cite{Olsen:2015zcy,Brambilla:2019esw,ParticleDataGroup:2024cfk}. These discoveries have spurred extensive studies aimed at deciphering the internal structure of such states. Among the various theoretical approaches, effective field theories like potential Non-Relativistic QCD (pNRQCD) \cite{Pineda:1997bj,Brambilla:1999xf} and the Born-Oppenheimer Effective Field Theory (BOEFT) \cite{Juge:1999ie,Braaten:2014qka,Berwein:2015vca,Oncala:2017hop,Brambilla:2018pyn,Brambilla:2019jfi,TarrusCastella:2017lex,Soto:2020xpm,Berwein:2024ztx} offer systematic frameworks for describing heavy quarkonium and hybrid states. In these approaches, the hybrid spectrum, decay rates, and transition amplitudes (including both spin-conserved and spin-flip processes) can be computed with inputs from lattice QCD studies \cite{Braaten:2014ita,Oncala:2017hop,TarrusCastella:2021pld,Brambilla:2022hhi,Bruschini:2024fyj,Braaten:2024stn}. 

In this work, we update the hybrid spectrum presented in ref. \cite{Oncala:2016wlm,Oncala:2017hop} for charmonium and bottomonium
incorporating recent lattice data of refs. \cite{Capitani:2018rox,Schlosser:2021wnr,Alasiri:2024nue}.
We next focus on the impact of the new data in the decay amplitudes of heavy hybrid mesons into conventional quarkonia. 
We recalculate the spin-conserved transitions presented in \cite{Oncala:2016wlm,Oncala:2017hop}, and extend the calculation to spin-flip transitions in \cite{TarrusCastella:2021pld,Brambilla:2022hhi}. 
The latter, despite being suppressed by powers of the heavy-quark mass, can be phenomenologically significant. We perform a comprehensive error analysis that includes: uncertainties in the masses of the bound states; higher-order corrections in the strong coupling; higher-order terms in the multipole expansion; relativistic corrections; and the replacement of the perturbative potential by the full (confining) potential. The latter contribution turns out to be significantly high. Finally, we compare our theoretical predictions with other approaches, and with experimental candidates listed in the Particle Data Group (PDG) \cite{ParticleDataGroup:2024cfk}, in order to identify potential hybrid states among the observed resonances.\\

The paper is organized as follows. In Section \ref{sec:EFTs}, we outline the two relevant effective field theories needed in our developments, the weak coupling pNRQCD and the BOEFT. In Section \ref{sec:trans_rates}, we describe the amplitud for hybrid to quarkonium transitions. Section \ref{sec:spin_con} details the computation of decay widths for spin-conserved transitions, while Section \ref{sec:spin_flip} addresses spin-flip transitions. In Section \ref{sec:comparision}, we compare our results with experimental data and discuss their implications for identifying exotic hadrons. Finally, Section \ref{sec:conclusion} summarizes our results and conclusions.

\begin{table}
\begin{minipage}{0.5 \textwidth}
\resizebox{\textwidth}{!}{  
	\begin{tabular}{|c|c|c|c|c|c|c|}
		\hline
        \rowcolor{cyan!20}
		&     &            &        &      ${\mathbb S}=0$      &      ${\mathbb S}=1$      &                             \\
        \rowcolor{cyan!20}
		$nL_J$ & w-f & $M_{c\bar{c}}$ & $M_{c\bar{c}g}$ & ${\cal J}^{PC}$ & ${\cal J}^{PC}$ & $\Lambda^{\epsilon}_{\eta}$ \\ \hline 
		$1s$ & S & 3068  &  & $0^{-+}$ & $1^{--}$ & $\Sigma_g^+$ \\ 
		$2s$ & S & 3674  &  & $0^{-+}$ & $1^{--}$ & $\Sigma_g^+$ \\ 
		$3s$ & S & 4149  &  & $0^{-+}$ & $1^{--}$ & $\Sigma_g^+$ \\ 
		$1p_0$ & $P^+$ &  & 4455   & $0^{++}$ & $1^{+-}$ & $\Sigma_u^-$ \\ 
		$4s$ & S & 4562  &  & $0^{-+}$ & $1^{--}$ & $\Sigma_g^+$ \\ 
		$2p_0$ & $P^+$ &  & 4917 & $0^{++}$ & $1^{+-}$ & $\Sigma_u^-$ \\ 
        $5s$ & S & 4937  &  & $0^{-+}$ & $1^{--}$ & $\Sigma_g^+$ \\ 
		$3p_0$ & $P^+$ &  & 5315  & $0^{++}$ & $1^{+-}$ & $\Sigma_u^-$ \\ 
		$4p_0$ & $P^+$ &  & 5650 & $0^{++}$ & $1^{+-}$ & $\Sigma_u^-$ \\ \hline
		$1p$ & S & 3457  &  & $1^{+-}$ & $(0,1,2)^{++}$ & $\Sigma_g^+$ \\ 
		$2p$ & S & 3958  &  & $1^{+-}$ & $(0,1,2)^{++}$ & $\Sigma_g^+$ \\ 
		$1(s/d)_1$ & $P^{+-}$ &  & 4028  & $1^{--}$ & $(0,1,2)^{-+}$ & $\Pi_u\Sigma_u^-$ \\ 
		$1p_1$ & $P^0$ &  & 4171  & $1^{++}$ & $(0,1,2)^{+-}$ & $\Pi_u$ \\ 
		$2(s/d)_1$ & $P^{+-}$ &  & 4394  & $1^{--}$ & $(0,1,2)^{-+}$ & $\Pi_u\Sigma_u^-$ \\ 
		$3p$ & S & 4388  &  & $1^{+-}$ & $(0,1,2)^{++}$ & $\Sigma_g^+$ \\
		$2p_1$ & $P^0$ &  & 4556  & $1^{++}$ & $(0,1,2)^{+-}$ & $\Pi_u$ \\ 
		$3(s/d)_1$ & $P^{+-}$ &  & 4678  & $1^{--}$ & $(0,1,2)^{-+}$ & $\Pi_u\Sigma_u^-$ \\ 
		$4(s/d)_1$ & $P^{+-}$ &  & 4755  & $1^{--}$ & $(0,1,2)^{-+}$ & $\Pi_u\Sigma_u^-$ \\ 
		$4p$ & S & 4774  &  & $1^{+-}$ & $(0,1,2)^{++}$ & $\Sigma_g^+$ \\ 
		$3p_1$ & $P^0$ &  & 4912  & $1^{++}$ & $(0,1,2)^{+-}$ & $\Pi_u$ \\ 
		$5(s/d)_1$ & $P^{+-}$ &  & 5087  & $1^{--}$ & $(0,1,2)^{-+}$ & $\Pi_u\Sigma_u^-$ \\ 
		$5p$ & S & 5130  &  & $1^{+-}$ & $(0,1,2)^{++}$ & $\Sigma_g^+$ \\ \hline
		$1d$ & S & 3762  &  & $2^{-+}$ & $(1,2,3)^{--}$ & $\Sigma_g^+$ \\ 
		$2d$ & S & 4209  &  & $2^{-+}$ & $(1,2,3)^{--}$ & $\Sigma_g^+$ \\ 
		$1(p/f)_2$ & $P^{+-}$ &  & 4245  & $2^{++}$ & $(1,2,3)^{+-}$ & $\Pi_u\Sigma_u^-$ \\ 
		$1d_2$ & $P^0$ &  & 4369  & $2^{--}$ & $(1,2,3)^{-+}$ & $\Pi_u$ \\ 
		$2(p/f)_2$ & $P^{+-}$ &  & 4601  & $2^{++}$ & $(1,2,3)^{+-}$ & $\Pi_u\Sigma_u^-$ \\ 
		$3d$ & S & 4608  &  & $2^{-+}$ & $(1,2,3)^{--}$ & $\Sigma_g^+$ \\ 
		$2d_2$ & $P^0$ &  & 4739 & $2^{--}$ & $(1,2,3)^{-+}$ & $\Pi_u$ \\ 
		$3(p/f)_2$ & $P^{+-}$ &  & 4892  & $2^{++}$ & $(1,2,3)^{+-}$ & $\Pi_u\Sigma_u^-$ \\ 
		$4d$ & S & 4974  &  & $2^{-+}$ & $(1,2,3)^{--}$ & $\Sigma_g^+$ \\ 
		$4(p/f)_2$ & $P^{+-}$ &  & 4945  & $2^{++}$ & $(1,2,3)^{+-}$ & $\Pi_u\Sigma_u^-$ \\ 
		$3d_2$ & $P^0$ &  & 5083  & $2^{--}$ & $(1,2,3)^{-+}$ & $\Pi_u$ \\ \hline
        $1f$ & S & 4029  &  & $3^{+-}$ & $(2,3,4)^{++}$ & $\Sigma_g^+$ \\ 
        $2f$ & S & 4441  &  & $3^{+-}$ & $(2,3,4)^{++}$ & $\Sigma_g^+$ \\ 
        $3f$ & S &  4817 &  & $3^{+-}$ & $(2,3,4)^{++}$ & $\Sigma_g^+$ \\ 
        $4f$ & S & 5166  &  & $3^{+-}$ & $(2,3,4)^{++}$ & $\Sigma_g^+$ \\  \hline
        $1g$ & S & 4272  &  & $4^{-+}$ & $(3,4,5)^{--}$ & $\Sigma_g^+$ \\ 
        $2g$ & S & 4658  &  & $4^{-+}$ & $(3,4,5)^{--}$ & $\Sigma_g^+$ \\ 
        $3g$ & S & 5015 &  & $4^{-+}$ & $(3,4,5)^{--}$ & $\Sigma_g^+$ \\ \hline
	\end{tabular}}
	\subcaption{ \small Charmonium (S) and hybrid ($P^{+-0}$)  spectrum computed with 
	$m_c=1.47$ GeV.
	}
	\label{cEspectrum}
    \end{minipage}
\hfill
    \begin{minipage}{0.5\linewidth}
		\centering
        \resizebox{\textwidth}{!}{  
		\begin{tabular}{|c|c|c|c|c|c|c|}
			\hline 
            \rowcolor{cyan!20}
			&     &            &        &  ${\mathbb S}=0$ & ${\mathbb S}=1$ & \\
            \rowcolor{cyan!20}
			$nL_J$ & w-f & $M_{b\bar{b}}$ & $M_{b\bar{b}g}$ & ${\cal J}^{PC}$ & ${\cal J}^{PC}$ & $\Lambda^{\epsilon}_{\eta}$ \\ \hline
			$1s$ & S & 9551 &  & $0^{-+}$ & $1^{--}$ & $\Sigma_g^+$ \\ 
			$2s$ & S & 10017 &  & $0^{-+}$ & $1^{--}$ & $\Sigma_g^+$ \\ 
			$3s$ & S & 10355 &  & $0^{-+}$ & $1^{--}$ & $\Sigma_g^+$ \\ 
			$4s$ & S & 10643 &  & $0^{-+}$ & $1^{--}$ & $\Sigma_g^+$ \\ 
            $5s$ & S & 10901 &  & $0^{-+}$ & $1^{--}$ & $\Sigma_g^+$ \\ 
			$1p_0$ & $P^+$ &  & 10977 & $0^{++}$ & $1^{+-}$ & $\Sigma_u^-$ \\ 
            $6s$ & S & 11139 &  & $0^{-+}$ & $1^{--}$ & $\Sigma_g^+$ \\ 
			$2p_0$ & $P^+$ &  & 11261 & $0^{++}$ & $1^{+-}$ & $\Sigma_u^-$ \\ 
			$3p_0$ & $P^+$ &  & 11525 & $0^{++}$ & $1^{+-}$ & $\Sigma_u^-$ \\ 
			$4p_0$ & $P^+$ &  & 11782 & $0^{++}$ & $1^{+-}$ & $\Sigma_u^-$ \\ \hline
			$1p$ & S & 9879  &  & $1^{+-}$ & $(0,1,2)^{++}$ & $\Sigma_g^+$ \\ 
			$2p$ & S & 10234 &  & $1^{+-}$ & $(0,1,2)^{++}$ & $\Sigma_g^+$ \\ 
			$3p$ & S & 10532 &  & $1^{+-}$ & $(0,1,2)^{++}$ & $\Sigma_g^+$ \\ 
			$1(s/d)_1$ & $P^{+-}$ &  & 10704 & $1^{--}$ & $(0,1,2)^{-+}$ & $\Pi_u\Sigma_u^-$ \\ 
			$1p_1$ & $P^0$ &  & 10772 & $1^{++}$ & $(0,1,2)^{+-}$ & $\Pi_u$ \\ 
			$4p$ & S & 10798 &  & $1^{+-}$ & $(0,1,2)^{++}$ & $\Sigma_g^+$ \\ 
			$2(s/d)_1$ & $P^{+-}$ &  & 10905 & $1^{--}$ & $(0,1,2)^{-+}$ & $\Pi_u\Sigma_u^-$ \\ 
			$2p_1$ & $P^0$ &  & 10995 & $1^{++}$ & $(0,1,2)^{+-}$ & $\Pi_u$ \\ 
			$5p$ & S & 11041 &  & $1^{+-}$ & $(0,1,2)^{++}$ & $\Sigma_g^+$ \\ 
			$3(s/d)_1$ & $P^{+-}$ &  & 11103 & $1^{--}$ & $(0,1,2)^{-+}$ & $\Pi_u\Sigma_u^-$ \\ 
			$4(s/d)_1$ & $P^{+-}$ &  & 11131 & $1^{--}$ & $(0,1,2)^{-+}$ & $\Pi_u\Sigma_u^-$ \\ 
			$3p_1$ & $P^0$ &  & 11209 & $1^{++}$ & $(0,1,2)^{+-}$ & $\Pi_u$ \\
			$6p$ & S & 11268 &  & $1^{+-}$ & $(0,1,2)^{++}$ & $\Sigma_g^+$ \\ 
			$5(s/d)_1$ & $P^{+-}$ &  & 11318 & $1^{--}$ & $(0,1,2)^{-+}$ & $\Pi_u\Sigma_u^-$ \\ \hline
			$1d$ & S & 10106 &  & $2^{-+}$ & $(1,2,3)^{--}$ & $\Sigma_g^+$ \\ 
			$2d$ & S & 10416  &  & $2^{-+}$ & $(1,2,3)^{--}$ & $\Sigma_g^+$ \\ 
			$3d$ & S & 10689  &  & $2^{-+}$ & $(1,2,3)^{--}$ & $\Sigma_g^+$ \\
			$1(p/f)_2$ & $P^{+-}$ &  & 10823  & $2^{++}$ & $(1,2,3)^{+-}$ & $\Pi_u\Sigma_u^-$ \\ 
			$1d_2$ & $P^0$ &  & 10880 & $2^{--}$ & $(1,2,3)^{-+}$ & $\Pi_u$ \\ 
			$4d$ & S & 10939 &  & $2^{-+}$ & $(1,2,3)^{--}$ & $\Sigma_g^+$ \\ 
			$2(p/f)_2$ & $P^{+-}$ &  & 11023  & $2^{++}$ & $(1,2,3)^{+-}$ & $\Pi_u\Sigma_u^-$ \\ 
			$2d_2$ & $P^0$ &  & 11100  & $2^{--}$ & $(1,2,3)^{-+}$ & $\Pi_u$ \\ 
			$5d$ & S & 11171 &  & $2^{-+}$ & $(1,2,3)^{--}$ & $\Sigma_g^+$ \\ 
			$3(p/f)_2$ & $P^{+-}$ &  & 11220  & $2^{++}$ & $(1,2,3)^{+-}$ & $\Pi_u\Sigma_u^-$ \\ 
			$3d_2$ & $P^0$ &  & 11310  & $2^{--}$ & $(1,2,3)^{-+}$ & $\Pi_u$ \\ 
			$4(p/f)_2$ & $P^{+-}$ &  & 11258  & $2^{++}$ & $(1,2,3)^{+-}$ & $\Pi_u\Sigma_u^-$ \\ \hline
            $1f$ & S &  10297 &  & $3^{+-}$ & $(2,3,4)^{++}$ & $\Sigma_g^+$ \\ 
            $2f$ & S & 10579  &  & $3^{+-}$ & $(2,3,4)^{++}$ & $\Sigma_g^+$ \\ 
            $3f$ & S & 10835  &  & $3^{+-}$ & $(2,3,4)^{++}$ & $\Sigma_g^+$ \\ 
            $4f$ & S & 11072  &  & $3^{+-}$ & $(2,3,4)^{++}$ & $\Sigma_g^+$ \\  \hline
            $1g$ & S & 10624  &  & $4^{-+}$ & $(3,4,5)^{--}$ & $\Sigma_g^+$ \\ 
            $2g$ & S & 10873  &  & $4^{-+}$ & $(3,4,5)^{--}$ & $\Sigma_g^+$ \\ 
            $3g$ & S & 11104 &  & $4^{-+}$ & $(3,4,5)^{--}$ & $\Sigma_g^+$ \\ \hline
		\end{tabular}}
		\subcaption{ \small Bottomonium (S) and hybrid ($P^{+-0}$)  spectrum computed with $m_b=4.88$ GeV.\label{bEspectrum}}
    \end{minipage}
    \caption{ \footnotesize  \label{fig:hybrid_spectrum}
		Quarkonium (S) and Hybrid quarkonium ($P^{+-0}$) spectrum. States which only differ by the heavy quark spin $ ({\mathbb S}=0,1)$ are degenerated. Units are in MeV. $n$ is the principal quantum number, $L$ the orbital angular momentum of the heavy quarks, $J$ is $L$ plus the total angular momentum of the gluons, $\mathbb S$ the spin of the heavy quarks and ${\cal J}$ the total angular momentum. For quarkonium, $J$ coincides with $L$, and it is not displayed. The last column shows the relevant static potentials for each state. The $(s/d)_1$, $p_1$, $p_0$, $(p/f)_2$ and $d_2$ states are usually named $H_1$, $H_2$, $H_3$, $H_4$ and $H_5$ respectively \cite{Juge:1999ie,Berwein:2015vca,Brambilla:2022hhi}. The spectrum is an update of {\cite{Oncala:2017hop
        } computed using the static potentials \cite{
        Alasiri:2024nue}}. We shall associate a precision of about $110$ MeV for charmonium and $33$ MeV for bottomonium masses, as we discuss in Appendix \ref{sec:uncertainity_mass}.}
\end{table}

\begin{figure}[H]
    \centering
    \includegraphics[width=0.94\linewidth]{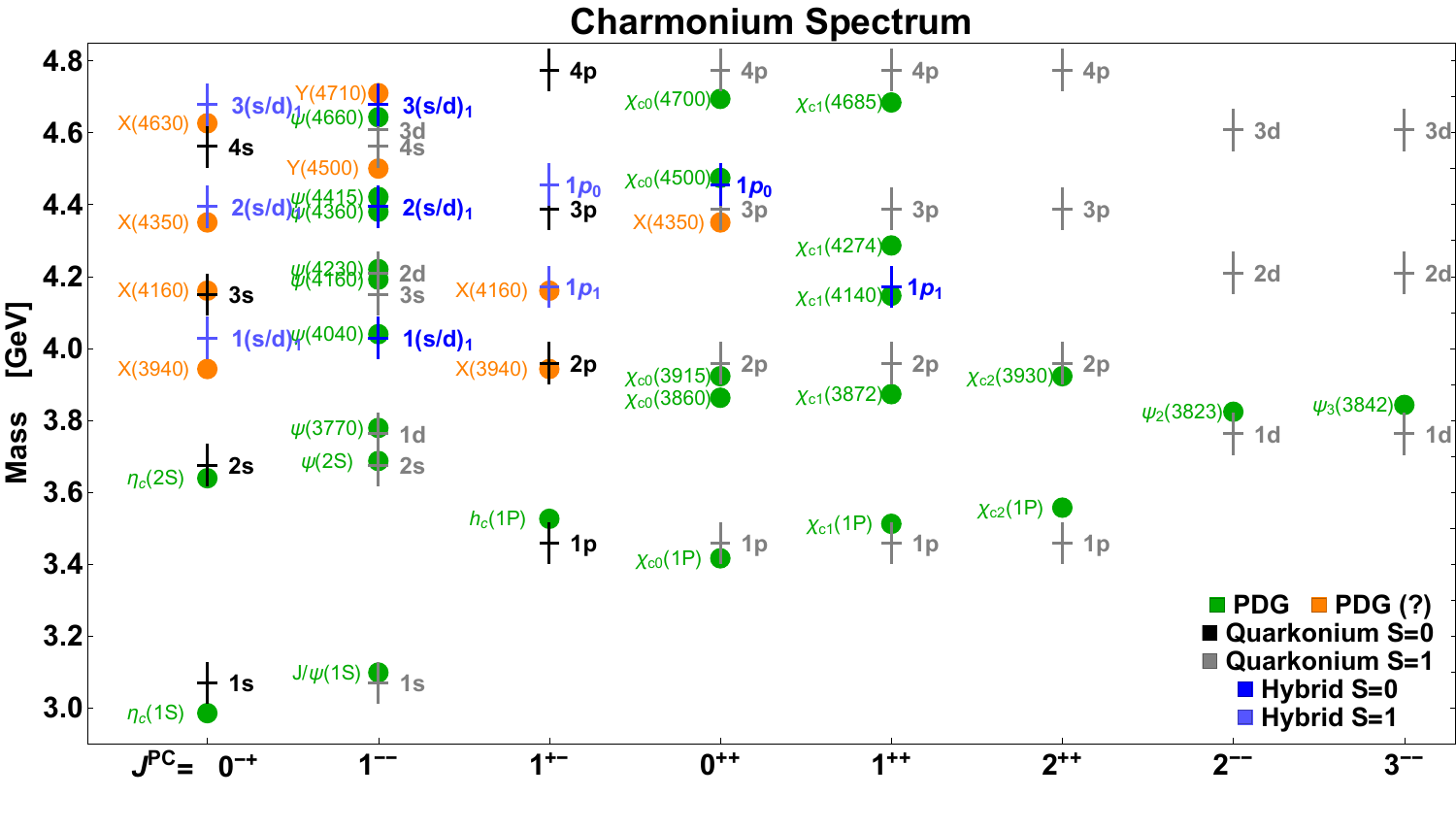}
    \includegraphics[width=0.94\linewidth]{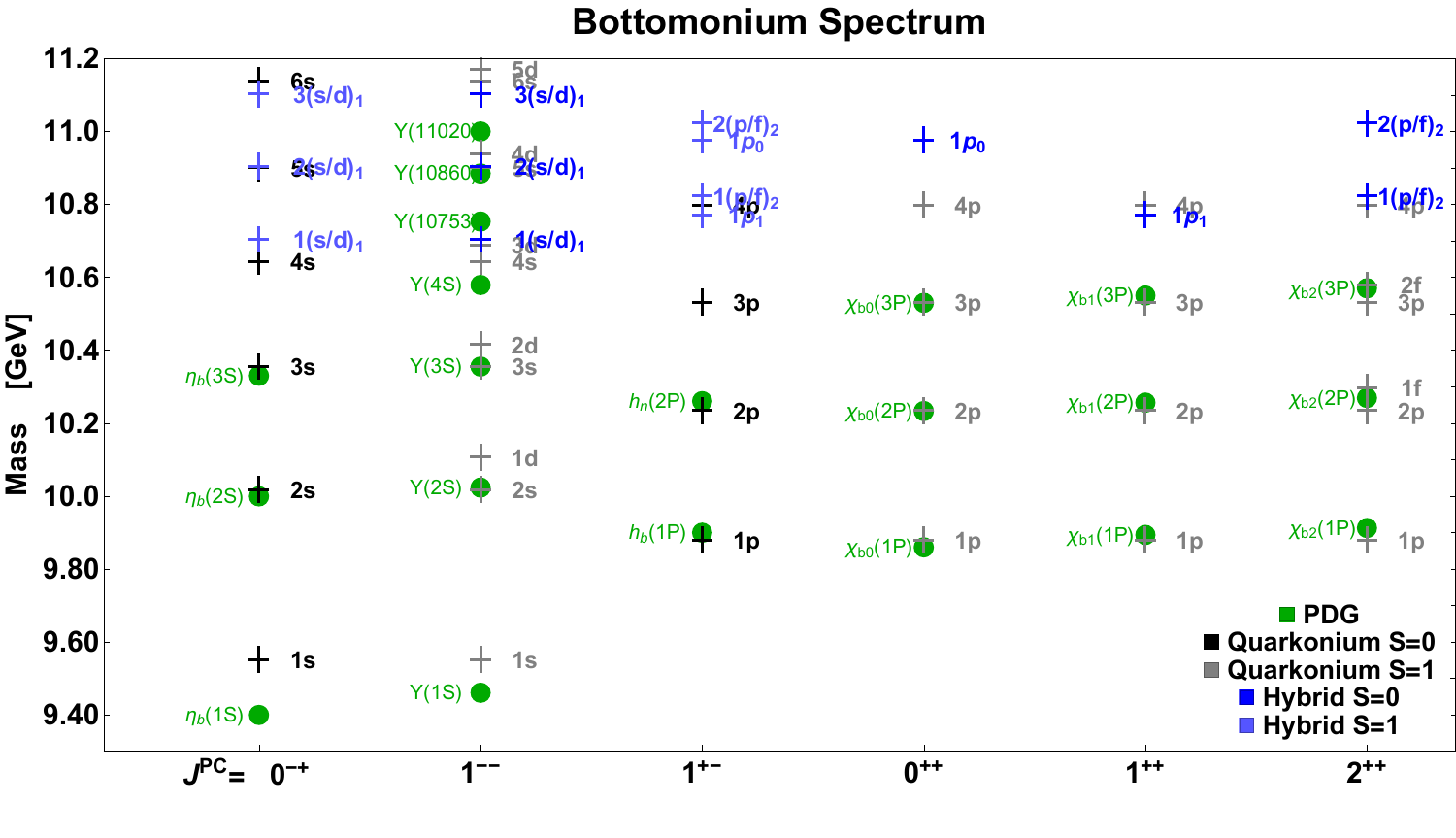}
    \caption{\footnotesize Spectrum of charmonium (up) and bottomonium (down) from  Table \ref{fig:hybrid_spectrum} {put on top of}  the experimental XYZ mesons reported by the PDG. We show in green the states with { known $J^{PC}$  and in orange the states with unknown or partially known $J^{PC}$}. We show an error of $\pm 110$ MeV ($\pm 33$ MeV) for charmonium (bottomonium) { according to} our expected precision. The $(s/d)_1$, $p_1$, $p_0$, $(p/f)_2$ and $d_2$ states are {usually named $H_1$, $H_2$, $H_3$, $H_4$ and $H_5$ respectively} \cite{Juge:1999ie,Berwein:2015vca, Brambilla:2022hhi}.  Our identifications are summarized in Tables \ref{tab:hybrids},\ref{tab:charmonium0}, and \ref{tab:hybrids0}.}
    \label{fig:spectrumC}
\end{figure}

\section{Effective Field Theories \label{sec:EFTs}}

We briefly describe here the two effective field theories that we need for our developments.

\subsection{Weak coupling pNRQCD}

pNRQCD is a non-relativistic effective field theory that describes quark-antiquark ($Q$-$\bar Q$) pairs at small relative {distance} (see \cite{RevModPhys.77.1423,Pineda:2011dg} for reviews). It is obtained from NRQCD \cite{Thacker:1990bm,Bodwin:1994jh} after integrating out the energy scale $1/r\sim m_Q v$, where $r=\vert\mathbf{r}\vert$ and $m_Q v$, $v\ll 1$, are the typical distance and the typical relative momentum between quark and antiquark, and $m_Q$ the heavy quark mass. If $1/r\gg \lQ$, the typical hadronic scale, the integration can be carried out in perturbation theory of $\als (1/r)$ at any order of the $1/m_Q$ expansion. The degrees of freedom of pNRQCD are: $Q$-$\bar Q$ pair with energy much smaller than $1/r$ and relative momentum $\sim m_Q v$ and ultrasoft gluons and light quarks, with energy and momentum much smaller than $1/r$. The $Q$-$\bar Q$ pair can be decomposed into a singlet field $S({\bf R},{\bf r},t)$ and an octet field $O({\bf R},{\bf r},t)$, in relation to colour gauge transformation  with respect to the centre-of-mass coordinate, ${\bf R}$. The ultrasoft fields are evaluated at ${\bf R}$, i.e.  $A_\mu = A_\mu({\bf R},t)$, and hence they do not depend on ${\bf r}$. This is due to the fact that the typical size of the system ${\bf r}$ is small with respect to the typical hadronic distance $1/\lQ$, and hence gluon fields can be multipole expanded with respect to this variable. Consequently, the pNRQCD Lagrangian, ${\cal L}_{\rm pNRQCD}$, can be organized  as a sum of ${\cal L}_{\rm pNRQCD}^{(k,l)}$, where $k$ and $l$ denote the order in  the $1/m_Q$ and multipole ($r$) expansion respectively. The singlet and octet field dynamics is dictated by the singlet ($V_s$) and octet ($V_o$) potentials respectively (the latter also interacts with static ultrasoft gluons) at leading order (LO),
\be
{\cal L}_{\rm pNRQCD}^{(0,0)} =
{\rm Tr} \,\Biggl\{ {\rm S}^\dagger \left( i\partial_0 -\frac{{\bf p}^2}{m_Q}- V_s(r) + \dots  \right) {\rm S} 
+ {\rm O}^\dagger \left( iD_0 -\frac{{\bf p}^2}{m_Q}- V_o(r) + \dots  \right) {\rm O} \Biggr\}\,,
\label{pnrqcd}
\ee
where $iD_0 {\rm O} \equiv i \partial_0 {\rm O} - g [A_0({\bf R},t),{\rm O}]$, $V_s(r)\simeq-C_f\als/r$, $V_o(r)\simeq 1/2N_cr$, $C_f=(N_c^2-1)/2N_c$ and $N_c=3$ is the number of colors.
At next-to-leading order (NLO), singlet-octet interaction terms occur that will be relevant to us, namely,
\bea 
{\cal L}_{\rm pNRQCD}^{(0,1)}&=&  g V_A ( r) {\rm Tr} \left\{  {\rm O}^\dagger {\bf r} \cdot {\bf E} \,{\rm S}
+ {\rm S}^\dagger {\bf r} \cdot {\bf E} \,{\rm O} \right\}\label{Lagrangian_0}\\
{\cal L}_{\rm pNRQCD}^{(1,0)}&=& g\frac{c_F}{m_Q}{\rm Tr}
    \Biggl\{
    {\rm S}^\dagger({\bf s}_1 -{\bf s}_2)\cdot {\bf B} {\rm O}+
    {\rm O}^\dagger({\bf s}_1 -{\bf s}_2)\cdot \ {\bf B} {\rm S}
    \Biggr\}\label{Lagrangian_1}
\eea
The trace is over the spin and the color indices, $V_A(r)=1+\mathcal{O}(\als^2)$ and $c_F=1+\mathcal{O}(\als)$ are matching coefficients. The latter is inherited from NRQCD. 
${\bf E}$ and ${\bf B}$ are the chromoelectric and the chromomagnetic fields.
The matrices ${\bf s}_1 = {\bm \sigma}_1/2$ and ${\bf s}_2 = {-{\bm\sigma}}_2^\mathrm{T}/2$ are the spin of the heavy quark and heavy antiquark, respectively. 

\subsection{BOEFT}

When $1/r\sim \lQ$ the integration of this scale cannot be carried out in perturbation theory. However, an effective theory can still be built. It was called strong coupling pNRQCD in ref. \cite{ Brambilla:1999xf}{.  Nowadays it is known as BOEFT, since the leading-order corresponds to the Born-Oppenheimer approximation \cite{Hasenfratz:1980jv,Horn:1977rq,Perantonis:1990dy,Juge:1999ie}}. For heavy quarkonium, at leading order in the $1/m_Q$ expansion, it reads,
\be
{\cal L}=S^\dagger\left( i\partial_0-\frac{\nabla^2}{m_Q}-V_{\Sigma_g^+}(r)\right)S
\label{cornell}
\ee
$S=S({\bf R},{\bf r},t)$. At short distances, $V_{\Sigma_g^+}(r)\simeq V_s(r)$, namely it must approach the singlet potential in Eq. \eqref{pnrqcd}. At distances $1/\lQ \lesssim r$, $V_{\Sigma_g^+}(r)$ must be evaluated non-perturbatively, for instance in lattice QCD.

Analogous Lagrangians can be built for states whose heavy quark content approaches the octet field in Eq. \eqref{pnrqcd} at short distances. A general construction at order $1/m_Q$ can be found in \cite{Soto:2020xpm,Berwein:2024ztx}. Here, we only need the LO Lagrangian for hybrids with $J^{PC}=1^{+-}$ light degrees of freedom,
\bea
\label{h}
&{\cal L}={\rm tr} \left( {H^i}^\dagger\left( \delta_{ij}i\partial_0-{h_H}_{ij}\right)H_j \right)\\
&{h_{H}}_{ij}\!=\!\left(\!-\!\frac{\nabla^2}{m_Q}\!+\!V_{\Sigma_u^-}(r)\right)\delta_{ij}\!+\!\left(\delta_{ij}\!-\!\hat{r}_{i} \hat{r}_{j}\right)\left[V_{\Pi_u}(r)\!-\!V_{\Sigma_u^-}(r)\right]\nonumber
\eea
$H^i=H^i({\bf R},{\bf r},t)$. At short distances, $V_{\Pi_u}(r)\simeq V_{\Sigma_u^-}(r)\simeq V_o(r)$ in Eq. \eqref{pnrqcd}. At distances $ 1/\lQ \lesssim  r$, $V_{\Sigma_u^-}(r)$ and $V_{\Pi_u}(r)$ must be evaluated non-perturbatively, for instance in lattice QCD.

{We have recalculated the quarkonium and hybrid spectrum provided in ref. \cite{Oncala:2017hop} with the $V_{\Sigma_g^+}(r)$, $V_{\Pi_u}(r)$ and $V_{\Sigma_u^-}(r)$ potentials given in ref. \cite{Alasiri:2024nue}. They are obtained from recent lattice data \cite{Capitani:2018rox,Schlosser:2021wnr} and account for finite lattice spacing corrections. The outcome is displayed in Table \ref{fig:hybrid_spectrum} and Figure \ref{fig:spectrumC}. The differences with respect to ref. \cite{Oncala:2017hop} fall within the expected errors.}

\section{Hybrid to Quarkonium transition rates}
\label{sec:trans_rates}
\begin{figure}[H]
		\centering 
		\includegraphics[width=0.35\textwidth]{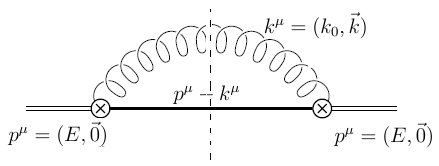}
		\caption{\small {Octet field self-energy diagram in pNRQCD}. Double line represents octet propagator (hybrid), while single lines represent singlet propagator (quarkonium). The curly line is a gluon of energy larger than $\lQ$, the vertices are chromoelectric dipole vertices Eq. \eqref{Lagrangian_0} for spin-conserved transitions or chromomagnetic dipole ones Eq. \eqref{Lagrangian_1} for spin-flip transitions. Gluons with energy similar to $\lQ$ are not displayed. They are necessary to make the initial and final state gauge invariant.}
		\label{decay}	
	\end{figure}

The lower lying heavy quarkonium states are amenable of a weak coupling treatment. Since the singlet and octet potentials are  attractive and repulsive respectively, a gap is expected between hybrids and lower lying quarkonium states. If we build an effective theory for hybrids, the lower-lying quarkonium states can be integrated out, which produces new octet-octet interaction terms. The relevant diagrams are depicted in Fig. \ref{decay}. Because the quarkonium states are lighter, this term will have an imaginary part that is related to the hybrid decay rate into quarkonium plus light hadrons. For spin-conserved transitions, this term reads, \cite{Oncala:2017hop},
\begin{equation}
	{\rm Im}\Delta V=-
	\frac{2}{3}\frac{\als T_F}{N_c} \sum_n r^i|S_n\rangle\langle S_n|r^i\, 
	(i\partial_t-E_n)^3\, ,
	\label{imdeltaVSC}
	\end{equation} 
    where we have already taken $V_A(r)\simeq 1$.
 For spin-flip transitions, which are suppressed by a factor $1/(m_Q r)^2$, we have \cite{Brambilla:2022hhi}
 \begin{equation}
	{\rm Im}\Delta V=-
	\frac{2}{3}\frac{\als T_F}{N_c}\frac{c_F^2}{m_Q^2}\sum_n \left(S_1^i-S_2^i\right)|S_n\rangle\langle S_n|\left(S_1^i-S_2^i\right)\, 
	(i\partial_t-E_n)^3\,,
	\label{imdeltaVSF}
	\end{equation} 
Note that from the formulas above one gets the hybrid to quarkonium inclusive decay width. However, since the contribution of each quarkonium state is easy to identify by picking up a single term in the sum, we are going to use it to calculate semi-inclusive decay rates. Let us then write
\be
{\rm Im}\Delta V=\sum_n {\rm Im}\Delta V_n\,,
\label{ImVn}
\ee
where ${\rm Im}\Delta V_n$ stands for the contribution of the $S_n$ quarkonium state to the inclusive decay width.  Hence, the semi-inclusive decay width to a single quarkonium state $S_n$ reads
\begin{equation}
    \Gamma(H\to S_{n}+X) = -2  \left<H|{\rm Im}\Delta V_n |H \right>\,,
    \label{defdecay}
\end{equation}
where we denote $\left|H\right>$  the decaying hybrid state. 
From now on we remove the subindex $n$ from the quarkonium state that we focus on and just call it $S$. In order to establish the validity of our calculation we must make sure that the conditions which lead to the diagram displayed in Fig. \ref{decay} hold. This are the expansion in $1/m_Q$ inherited from NRQCD, the expansion in $\mathbf{r}$ of pNRQCD, and the perturbative calculation of the diagram.
Concerning the latter, we must restrict ourselves to quarkonium and hybrid states for which
$
 \Delta E>>\Lambda_{QCD} \,,  
$
where $\Delta E$ is the mass difference between the hybrid and the quarkonium states and $\Lambda_{QCD}$ the typical hadronic scale. Otherwise the perturbative calculation does not hold. Concerning the expansion in $\mathbf{r}$,
we must require that
the emitted gluon cannot resolve the quark-antiquark pair distance, i.e., the matrix element of the heavy quark-antiquark distance $\bf r$ between the hybrid and the quarkonium states must be much smaller than $ 1/\Delta E$, i. e.
\begin{equation}\label{eq:multp_exp:spin-cons}
\left<H|{\bf r}|S_{}\right>\Delta E\ll 1 \,.
\end{equation}
Finally, concerning the $1/m_Q$ expansion, we require
\begin{equation} \label{eq:multp_exp:spin-flip}
\left<H|S_{}\right>\Delta E/m_Q\ll 1  \,,
\end{equation}
namely, that the non-relativistic expansion holds.
If these conditions are fulfilled, the transition $H \to S_{}$ can be treated in weakly-coupled pNRQCD, and the gluon at the scale $\Delta E$ is perturbative. In practice we are going to require that $\Delta E > 800$ MeV, 
and that Eq. (\ref{eq:multp_exp:spin-cons})  is fulfilled for spin-conserved transitions,  and Eq. (\ref{eq:multp_exp:spin-flip}) for spin-flip transitions (see Eqns. \eqref{eq:<r>_cons},\eqref{eq:<>0} and \eqref{eq:<>1} for the precise implementation). We will take into account errors due to higher order terms in the $\als$, $\mathbf{r}$ and $1/m_Q$ expansions (see Appendix \ref{sec:uncertainities}).

\section{Spin-conserved transitions}
\label{sec:spin_con}

The spin-conserved decay is computed following Eqns.
\eqref{imdeltaVSC},\eqref{ImVn} and \eqref{defdecay},
\begin{equation}
    \Gamma(H_{JM}\to S_{L'M'})=\frac{4 \alpha_s}{3 N_C} T_F \Delta E ^3 \left<H_{JM}^i|{\bf r}^{i'}|S_{L'M'}\right>^* \left<H_{JM}^i|{\bf r}^{i'}|S_{L'M'}\right>.
    \label{eq:Gamma}
\end{equation}
The matrix element is the expected value of ${\bf r}=r\ { \widehat{r}}$ computed between the wavefunctions of the hybrid initial state $\left<H_{JM}\right|$, which can be expressed in terms of vector spherical harmonics (VSH) as in Eq. \eqref{H}, and the quarkonium final $\left|S_{L'M'}\right>$ states, that is written with the usual scalar spherical harmonics (SH) form as in Eq. \eqref{S}, hence
\begin{equation}
    \left<H_{JM}^i|{\bf r}^{i'}|S_{L'M'}\right>=\int \frac{1}{r}\left[P_J^{+}(r) {Y^{*i}}_{JM}^{L=J+1} +P_J^{0}(r){{ Y}^{*i}}_{JM}^{L=J}+P_J^{-}(r) {{ Y}^{*i}}_{JM}^{L=|J-1|}\right] \  r \  {\widehat{r}^{i'}} \ \frac{R_{L'}(r)}{r}Y^{L'}_{M'} \ r^2 dr d\Omega.
    \label{eq:HrS}
\end{equation}
Using the radial projection of the SH in Eq. \eqref{rY} together with the expansion of the VHS in Eq. \eqref{expansion_VSH}, we shall get 6 terms from Eq. \eqref{eq:HrS}, which coefficients $\mathcal{C}_{JM \to L'M'}^{Ll'i i'}$ are the angular integrals, where the index $L=\{J+1,J,|J-1|\}$ runs for the initial hybrid components of the wave-function, while $l'=\{L'+1,|L'-1|\}$ for the final quarkonium components.
\begin{equation}\small
    \Gamma(H_{JM}\to S_{L'M'})=\frac{4 \alpha_s}{3 N_C} T_F \Delta E ^3  \ \left[ \sum_{L,l'}  \ \mathcal{C}^{Ll'i i'}_{JM\to L'M'}  \int P^L_J(r) \ r \ R_{L'}(r) \ dr\right] \left[ \sum_{L,l'}  \ \mathcal{C}^{Ll'i i'}_{JM\to L'M'}  \int P^L_J(r) \ r \ R_{L'}(r) \ dr\right]^\ast.
    \label{eq:full_decay1}
\end{equation}
The first  angular integral shall  read as:
\begin{align}
   \mathcal{C}^{L+ii'}_{JM \to L'M'}&\equiv -\sqrt{\frac{L'+1}{2L'+1}}\int_\Omega {{ Y}^{*i}}_{JM}^{L} {{ Y}^{i'}}^{L'+1}_{L'M'} \\
   &=-\sqrt{\frac{L'+1}{2L'+1}} \int_\Omega   \ \sum_{\mu} {Y^*}^{L}_ {M-\mu} C(L,1,J;M-\mu,\mu){\chi^*}^i_\mu \sum_{\mu'}Y^{L'+1}_ {M'-\mu'} C(L'+1,1,L';M'-\mu',\mu')\chi^{i'}_{\mu'}, \nonumber
\end{align}
which give rise to two deltas for angular and magnetic momentum conservation. By squaring in Eq. \eqref{eq:full_decay1}, we shall get $
({2\times 3})^2=36$ terms: 6 non-crossed and 30 crossed-terms. Each of them carries an angular {momentum} selection rule, and some conservation of the spinor indices, e.i. 
$
     \left[{\chi^*}^i_\mu \chi^{i'} _{\mu'}\right]^*\cdot\left[ {\chi^*}^i_\nu \chi^{i'} _{\nu'}\right]=
     \delta_{\mu}^{\nu}\delta_{\mu'}^{\nu'}.
$
\\

{\bf Non-crossed terms} shall survive given selection rule in the angular momentum conservation.
\begin{align}
[\mathcal{C}^{L+}_{JM\to L'M'}]^2\equiv {\mathcal{C}^{L+ii'\,*}_{JM\to L'M'}}\mathcal{C}^{L+ii'}_{JM\to L'M'} &=\frac{L'+1}{2L'+1}
\sum_{\mu \mu'} C^2(L,1,J;M-\mu,\mu)  C^2(L'+1,1,L';M-\mu,\mu') \ \delta^{L}_{L'+1} \delta_{M'-\mu'}^{M-\mu} ,
\label{C++}\\
[\mathcal{C}^{L-}_{JM\to L'M'}]^2\equiv {\mathcal{C}^{L-ii'\,*}_{JM\to L'M'}}\mathcal{C}^{L-ii'}_{JM\to L'M'}&=\frac{L'}{2L'+1}\sum_{\mu \mu'} C^2(L,1,J;M-\mu,\mu)  C^2(|L'-1|,1,L';M-\mu,\mu') \ \delta^{L}_{|L'-1|} \delta_{M'-\mu'}^{M-\mu}   .
\label{C+-}
\end{align}
 Notice that we keep explicitly the delta of angular momentum conservation to remember what channel is allowed in each case (for example  $\delta^{L=J+1}_{L'+1}$ is allowed for $J=L'$).
\begin{table}
    \centering
    \scalebox{0.75}{
    \begin{tabular}{|c|c|c c|c c|c c|}
         \hline
         \rowcolor[gray]{0.8} 
         \multicolumn{2}{|c|}{ }& \multicolumn{2}{|c|}{ $P^{L=J+1}_J $ }&\multicolumn{2}{|c|}{$P^{L=J}_J $ } &\multicolumn{2}{|c|}{ $P^{L=J-1}_J$}\\
         \hline
         \rowcolor[gray]{0.8} 
         $J$ &  $L'$ & $[\mathcal{C}^{++}_{JL'}]^2$ & $[\mathcal{C}^{+-}_{JL'}]^2$ & $[\mathcal{C}^{0+}_{JL'}]^2$ & $[\mathcal{C}^{0-}_{JL'}]^2$ & $[\mathcal{C}^{-+}_{JL'}]^2$ & $[\mathcal{C}^{--}_{JL'}]^2$ \\ \hline
        0 & 0 & 1/3 & -   & - & - & - & - \\ 
        0 & 1 & x   & x   & - & - & - & -\\ 
        0 & 2 & x   & 2/3 & - & - & - & -\\ \hline 
        1 & 0 & x   & -   & 1/3 & - & x & - \\ 
        1 & 1 & 2/5 & x   & x & x & x & 1\\ 
        1 & 2 & x   & x   & x & 2/3 & x & x\\
        1 & 3 & x   & 3/5 & x & x & x & x\\ \hline  
    \end{tabular}}
    \caption{\small Selection rules and numerical coefficients for {spin-conserved} hybrid to quarkonium transitions. We mark with ``x" the {transitions that are not allowed}, while we mark with ``-" the cases that do not exist.}
    \label{tab:selection_rules}
\end{table}

{\bf Crossed terms} vanish, they shall generally arise from the structure:
\begin{align}
\label{crossC}
[\mathcal{C}^{L+ ii'}_{JM\to L'M'}]^*[\mathcal{C}^{\tilde{L}- ii'}_{JM\to L'M'}]=
\frac{\sqrt{(L'+1)L'}}{2L'+1}
\sum_{\mu \mu'} &C(L,1,J;M-\mu,\mu)  C(L'+1,1,L';M-\mu,\mu') \\
&C(\tilde{L},1,J;M-\mu,\mu)  C(|L'-1|,1,L';M-\mu,\mu') \ \delta^{L}_{L'+1}  \delta^{\tilde{L}}_{|L'-1|} \delta_{M'-\mu'}^{M-\mu}, \nonumber
\end{align}
where $L$ and $\tilde{L}$ can take any value in $\{J+1, J, |J-1|\}$.
\begin{itemize}
    \item For $J=0$, $L$ and $\tilde{L}$ can only take the value $L=\tilde{L}=J+1=1$, but this leads to $L'=0$ given the delta of angular momentum conservation, and this term vanishes for the prefactor in Eq. \eqref{crossC}.
    \item For $J=1$, the accessible cross-term comes from $L'=0,1$.
    Nevertheless, these terms also vanish: the case $L'=0$  due to the prefactor in Eq. \eqref{crossC}, while $L'=1$ when doing the Clebsch-Gordan sums.
\end{itemize}
{\bf Spin average}: for $\mathbb{S}=0$ hybrids, the initial hybrid state is characterized by $JM{L}$ while the final quarkonium state by $L'M'$, where $JM{L}$ and $L'M'$ are fixed in each transition. If one does not know the $M$ of the initial state or does not measure the $M'$ of the final state, the spin average of the squared amplitudes is done by:
\begin{equation}
   \left[{\cal C}_{J L'}^{Ll}\right]^2\equiv \frac{1}{2J+1}\sum_{M=-J}^{J}\sum_{M'=-L'}^{L'} \left[{\cal C}^{Ll}_{JM\to L'M'}  \right]^2 \
   \label{coefficients}
\end{equation}
The coefficients of Eq. \eqref{coefficients} can be computed for each combination of $J{L}$ and $L'$ and are given in  Table \ref{tab:selection_rules}. Putting all together, the general expression for the spin-conserved transition reads as:
\begin{equation}
    \Gamma(H_J\to S_{L'})=\frac{4 \alpha_s}{3 N_C} T_F \Delta E ^3 \sum_{Ll} \ [\mathcal{C}^{Ll}_{JL'}]^2 \left[ \int P^L_J(r) \ r \ R_{L'}(r) \ dr\right]^2\,.
    \label{eq:full_decay}
\end{equation}
In the following sections we apply this formula to the hybrid states $p_0$, $p_1$ and $(s/d)_1$. The results
can be found in Table \ref{tab:decay_conserved}, and the particular values for $\mathcal{C}^{Ll}_{JL'}$ in Table \ref{tab:selection_rules}. In order to estimate the reliability of our calculation, we use
in Table \ref{tab:decay_conserved},
\begin{equation}\label{eq:<r>_cons}
{|\langle H_J|r|S_{L'}\rangle|} 
   = \  \sum_{Ll} \left| \mathcal{C}^{Ll}_{JL'} \int P^L_J(r) \ r \ R_{L'}(r) \ dr\right| 
   \,.
\end{equation}
The expressions for the $\mathbb{S}=1$ hybrids depend on the same integrals, but with different Clebsch-Gordan coefficients. The identified PDG states listed in Table \ref{tab:XYZ} with  $\mathbb{S}=1$,  which may decay through spin-conserved channels, are all subject to large corrections in $\Gamma$, especially due to the contributions arising from the difference between weak- and strong-coupling regimes. Further details on these specific transitions are provided in the supplemental material ~\cite{oncala_mesa_2025_hybrid}. 

\subsection{The decay of the hybrid states $p_0$ \label{sec:decayJ0}}

The hybrid states with $J=0$ and $L=J+1=1$ feels the potential $ \Sigma_u^-$. It can be studied with a decoupled Schroedinger equation, with the only wave-function  $P^+_{0}$, giving rise to the hybrid states $(np_0)$. It is a state $J^{PC}=0^{++}$ at spin 0, and  $J^{PC}=1^{+-}$ at spin 1.  Its decay to quarkonium has two possible channels $L'=0,2$ with wave-function $R_{0,2}$ labeled as $(ns)$ and $(nd)$. 
From Eq. \eqref{eq:full_decay} the expression for the decay width is:
\begin{align}
    \Gamma(p_0\to s)&= \frac{4 \alpha_s}{3 N_C} T_F \Delta E ^3 \times \frac{1}{3}  \left[ \int P^+_{0}(r) \ r \ R_{0}(r) \ dr\right]^2 ,\\
    \Gamma(p_0\to d)&= \frac{4 \alpha_s}{3 N_C} T_F \Delta E ^3 \times
    \frac{2}{3}  \left[ \int P^+_{0}(r) \ r \ R_{2}(r) \ dr\right]^2.
\end{align}

\subsection{The decay of the hybrid states $p_1$ \label{sec:decayJ10}}

The hybrid states with $J=1$ and $L=J=1$ feel the potential $\Pi_u$. It can be studied with a decoupled Schroedinger equation, with the only wave-function  $P^0_{1}$, giving rise to the hybrid states $(np_1)$. It is a state $J^{PC}=1^{++}$ at spin 0, and  $J^{PC}=(0,1,2)^{+-}$ at spin 1.  Its decay to quarkonium has two possible channels $L'=0,2$ with wave-function $R_{0,2}$ labeled as $(ns)$ and $(nd)$. From Eq. \eqref{eq:full_decay}, the expression for the decay width is:
\begin{align}
    \Gamma(p_1\to s)&= \frac{4 \alpha_s}{3 N_C} T_F \Delta E ^3 \times
    \frac{1}{3}  \left[ \int P^0_{1}(r) \ r \ R_{0}(r) \ dr\right]^2,\\
    \Gamma(p_1\to d)&= \frac{4 \alpha_s}{3 N_C} T_F \Delta E ^3 \times
    \ \frac{2}{3}  \left[ \int P^0_{1}(r) \ r \ R_{2}(r) \ dr\right]^2.  
\end{align}

\subsection{The decay of the hybrid states $(s/d)_1$ \label{sec:decayJ1pm}}
The hybrid states with $J=1$ and $L=J\pm1=0,2$ feel the potential $\Pi_u$ and $\Sigma_u^-$. It can be studied with a coupled Schroedinger equation, with the wave-functions  $P^{+-}_{1}$, giving rise to the hybrid states $n(s/d)_1$. It is a state $J^{PC}=1^{--}$ at spin 0, and  $J^{PC}=(0,1,2)^{-+}$ at spin 1.  Its decay to quarkonium has two possible channels $L'=1,3$ with wave-function $R_{1,3}$ labeled as $(np)$ and $(nf)$. From Eq. \eqref{eq:full_decay}, the expression for the decay width is:
\begin{align}
    \Gamma[(s/d)_1\to p]&= \frac{4 \alpha_s}{3 N_C} T_F \Delta E ^3 \times \left(
     \frac{2}{5}  \left[ \int P^+_{1}(r) \ r \ R_{1}(r) \ dr\right]^2+ 1  \left[ \int P^-_{1}(r) \ r \ R_{1}(r) \ dr\right]^2 \right),\\
    \Gamma[(s/d)_1\to f]&= \frac{4 \alpha_s}{3 N_C} T_F \Delta E ^3 \times \frac{3}{5}  \left[ \int P^+_{1}(r) \ r \ R_{3}(r) \ dr\right]^2.
\end{align}

\begin{table}
\centering
\scalebox{0.7}{
\begin{tabular}{|cc|cc|ccc|}
\hline
\rowcolor{cyan!20}
\multicolumn{2}{|c|}{Hybrid}    & \multicolumn{2}{c|}{Quarkonium}             & \multicolumn{3}{c|}{Decay}            \\
\rowcolor{cyan!20}
$nL_J$     & $E$   & $n'L'$ & $E'$  & $\Delta E$  &   $\langle H| r|S \rangle \Delta E$ &$\Gamma\pm \delta \Gamma$ \\ \hline
\multicolumn{7}{|c|}{\cellcolor{lightgray}\textbf{Charm}}     \\ \hline
$2p_0$     & 4917  & $1s$   & 3068  & 1849  &      0.54            & 37    $\pm 29$ \\
$3p_0$     & 5315  & $1s$   & 3068  & 2247 &       0.26            & 9     $\pm 6$ \\ 
\hline
$2p_1$     & 4556  & $1s$   & 3068  & 1488 &       0.70            & 54     $\pm 52$ \\
$3p_1$     & 4912  & $1s$   & 3068  & 1844 &       0.42            & 22     $\pm 18$ \\
\hline
\multicolumn{7}{|c|}{\cellcolor{lightgray}\textbf{Bottom}}  \\ \hline
$2p_0$     & 11261 & $1s$   & 9551  & 1710  &      0.46            & 25    $\pm 13$ \\
$3p_0$     & 11525 & $1s$   & 9551  & 1974 &       0.29            & 11     $\pm 5$ \\
\hline
$1p_1$     & 10772 & $1s$   & 9551  & 1221  &      0.72            & 52     $\pm 38$ \\
$2p_1$     & 10995 & $1s$   & 9551  & 1444  &       0.55            & 32     $\pm 20$ \\
$3p_1$     & 11209 & $1s$   & 9551  & 1658  &      0.39            & 18     $\pm 10$ \\
\hline
\end{tabular}}
\caption{
\small \label{tab:decay_conserved} Table of spin-conserved transitions for hybrid charmonium and bottomonium. Units in MeV. In the table we only show the transitions that satisfy constraints in $\Delta E>800$ MeV and  $\langle{H| r|S}\rangle\Delta E<1$.
Last column includes theoretical error associated to higher
orders in $\alpha_s$,  the multipole expansion, the difference between weak and strong coupling regimes, relativistic corrections, as well as the imprecision in $\Delta E$, see Appendix \ref{sec:uncertainities}.}
\end{table}

\section{Spin-flip transitions}
\label{sec:spin_flip}
The chromomagnetic-dipole interaction in the Lagrangian Eq. \eqref{Lagrangian_1} is responsible for spin-flip decays of hybrid to quarkonium (spin 0 hybrid decaying to spin 1 quarkonium and vice versa). Spin-flip transition widths are, in principle, suppressed by powers of the heavy-quark mass due to the heavy-quark spin symmetry. They turn out to be smaller than spin-conserved transition widths because in any non-relativistic system $r\gg 1/m_Q$. How smaller they are depends on the relative
size of $\left<S|H\right>/m_Q$ with respect to $\left<S|{\bf r}|H\right>$,
and on the size of the energy gap between the hybrid and the quarkonium state. The spin-flip decay is given by
\begin{align}
    \Gamma(H_{J}\to S_{L'})&=\frac{4 \alpha_s}{3 N_C} T_F \Delta E ^3 \frac{\left<H_J|s_1^j-s_2^j|S_{L'}\right>^*}{m_Q} \frac{\left<H_J|s_1^j-s_2^j|S_{L'}\right>}{m_Q},
    \label{eq:Gamma_spinF}
\end{align}
where $s_1^j$ and $s_2^j$ are the spin vectors of the heavy quark and antiquark. In Appendix \ref{app:spin_fliped_element}, we expand the spin-flip matrix element. The {only non-vanishing results correspond to spin 0 hybrids decaying} into spin 1 quarkonium as:
\begin{align}
    \Gamma(H_{J}^{\text{spin} 0}\to S_{L'}^{\text{spin}1})=\frac{4 \alpha_s}{3 N_C} T_F \frac{\Delta E ^3}{m_Q^2} [S^{k*}H^n]^*[S^{k*}H^n],
    \label{eq:Gamma_spinF1}
\end{align}
and spin 1 hybrids decaying into spin 0 quarkonium as:
\begin{align}
    \Gamma(H_{J}^{\text{spin}1}\to S_{L'}^{\text{spin}0})=\frac{4 \alpha_s}{3 N_C} T_F \frac{\Delta E ^3}{m_Q^2} [S^*H^{nk}]^*[S^*H^{nk}],
    \label{eq:Gamma_spinF2}
\end{align}
where $k$ and $n$ are the vector/tensor indices of the wave-function.
\subsection{Spin 0 hybrid to spin 1 quarkonium}

The hybrid wave function for spin 0 is a vector wave function as in Eq. \eqref{H} and the spin 1 quarkonium wave function is also a vectorial wave function as in Eq. \eqref{S1}. The amplitude for the transition shall read as:
\begin{align}
    S^{k*}H^n= \int &\frac{1}{r}\left[P_J^{+}(r){ Y^n}_{JM}^{L=J+1} +P_J^{0}(r){ Y^n}_{JM}^{L=J}+P_J^{-}(r){ Y^n}_{JM}^{L=|J-1|}\right]  \frac{R_{L'}(r)}{r}{ {Y^k}^*}^{L'}_{J' M'} \ r^2 dr d\Omega,
    \label{eq:HsS}
\end{align}
The angular integral shall be:
\begin{equation}
    \mathcal{K}^{L \ nk}_{JM\to J'L'M'}\equiv  \int_\Omega { Y^n}_{JM}^{L} \cdot { {Y^k}^*}^{L'}_{J'M'} =\sum_{\mu\mu'} C(L,1,J;M-\mu,\mu)C(L',1,J';M'-\mu',\mu') \ \delta^{L}_{L'} \ \delta^{M-\mu}_{M'-\mu'} \ {\chi}^{\ n}_{\mu} {\chi}^{k \ *}_{\mu'}.
\end{equation}
By squaring, the non-crossed terms are:
\begin{align}
    [\mathcal{K}_{JM\to J'L'M'}^{L}]^2\equiv\mathcal{K}^{L \ nk \,\,{}^\ast}_{JM\to J'L'M'}\mathcal{K}^{L \ nk}_{JM\to J'L'M'}  &=\sum_{\mu\mu'} C^2(L,1,J;M-\mu,\mu)C^2(L,1,J';M-\mu,\mu') \delta_{L'}^{L} {\delta_{M'-\mu'}^{M-\mu}},
\end{align}
where $L$ can take the values $L=\{J+1,J,|J-1|\}$ for the initial hybrid components of the wave-function. 

\begin{table}[]
\centering
\scalebox{0.8}{
\begin{tabular}{|c|c|c|c|c|} \hline
\rowcolor[gray]{0.8} 
$J$ & $J'$ & $[\mathcal{K}_{J\to J'L'}^{L=J+1}]^2$ & $[\mathcal{K}_{J\to J'L'}^{L=J}]^2$ & $[\mathcal{K}_{J\to J'L'}^{L=J-1}]^2$ \\  \hline 
0   & 0    & 1/3    & -      & -    \\ 
0   & 1    & 1      & -      & -    \\  
0   & 2    & 5/3    & -      & -    \\ \hline
1   & 0    & x      & 1/3    & x    \\
1   & 1    & 3/5    & 1      & 3    \\ 
1   & 2    & 1      & 5/3    & x    \\ 
1   & 3    & 7/5    & x      & x    \\ \hline 
\multicolumn{5}{c}{
\hspace{1.5cm}$\delta_{L'}^{J+1}$ 
\hspace{1.3cm} $\delta_{L'}^{J}$
\hspace{1.3cm} $ \delta_{L'}^{|J-1|}$}
\end{tabular}}
\caption{\small \label{tab:selection_rules2} Table of selection rules for spin-flip transitions: Spin 0 hybrid to spin 1 quarkonium. Notice that for $J=0$ only exists the state $L=J+1=1$. We mark with ``x" the {transitions that are not allowed}, while we mark with ``-" the cases that do not exist.}
\end{table}
\textbf{Crossed terms:} The delta of angular momentum conservation does not allow the existences of any crossed term for $J>0$.
\begin{align}
    [\mathcal{K}_{JM\to J'L'M'}^{+\ nk}][\mathcal{K}_{JM\to J'L'M'}^{- \ nk}]\sim \delta^{J+1}_{L'}\delta^{|J-1|}_{L'}  =0.
\end{align}
{\bf Spin average}: the initial state in our transition is characterized by $JLM$ and the final state by $J'L'M'$, where $JLM$ and $J'L'M'$ are fixed in each transition. If one does not know the $M$ of the initial state or does not measure the $M'$ of the final state, the spin average of the squared amplitudes is done by:
\begin{equation}
   \left[{\cal K}_{J\to J'L'}^{L}\right]^2= \frac{1}{2J+1}\sum_{M=-J}^{J}{\sum_{M'=-J'}^{J'}} \left[{\cal K}^{L}_{JM\to J'L'M'}  \right]^2
   \label{coefficients2}
\end{equation}
The coefficients of Eq. \eqref{coefficients2} can be computed for each combination of $J$ and $J'L'$ and are given in  Table \ref{tab:selection_rules2}. Putting all together, the general expression for spin 0 hybrid to spin 1 quarkonium transition is:
\begin{equation}
    \Gamma(H_{J}^{\text{spin}0}\to S_{J'L'}^{\text{spin}1})=\frac{4 \alpha_s}{3 N_C} T_F \frac{\Delta E ^3}{m_Q^2} {\sum_L} \ \left[\mathcal{K}^{L}_{J\to J'L'} \ 
    \right]^2 \left[\int P^L_J(r)  \ R_{L'}(r) \ dr\right]^2 .
    \label{eq:full_decay2}
\end{equation}
To get the inclusive decay rate, we can sum all the final $J'$ states that contribute to the same $L'$ of the quarkonium wave-function. Notice that $L$ can take the values $J+1$, $J$ and $|J-1|$. 
{
In the following sections we apply this formula to the hybrid states $p_0$, $p_1$ and $(s/d)_1$. The results
can be found in Table \ref{tab:decay_flipped0} and the particular values for $\mathcal{K}^{L}_{J\to J'L'}$ in Table \ref{tab:selection_rules2}. In order to estimate the reliability of our calculation 
we use 
in Table \ref{tab:decay_flipped0}
\begin{equation}\label{eq:<>0}
|\langle H_{J}|S_{J'L'}\rangle |
  = \sum_L\left| \mathcal{K}^{L}_{J\to J'L'} 
   \int P^L_J(r)  \ R_{L'}(r) \ dr\right|
  \,.
\end{equation}
}

\subsubsection{The decay of the hybrid state $p_0$}
The hybrid state with $J=0$ and $L=J+1=1$ feels the potential $ \Sigma_u^-$. It can be studied with a decoupled Schroedinger equation, with the only wave-function  $P^{+}_{0}$ giving rise to the hybrid states $(np_0)$. It is a state $J^{PC}=0^{++}$ at spin 0.  Its spin-flip decay to quarkonium has one possible channel to $L'=1$ with wave-function $R_{1}$ labeled as $(np)$. The expression for the decay  {width} is:
\begin{align}
    \Gamma(p_0^{\text{spin}0}\to p^{\text{spin}1})=\frac{4 \alpha_s}{3 N_C} T_F \frac{\Delta E ^3}{m_Q^2} \times \left(\frac{1}{3}+1+\frac{5}{3}\right) \times \left[ \int P^+_0(r)  \ R_{1}(r) \ dr\right]^2  ,
    \label{eq:full_decayJ0}
\end{align}
In the parenthesis, the $1/3$ count for the final state $J'=0$, the $1$ for the final state $J'=1$ and $5/3$ for $J'=2$.

\subsubsection{The decay of the hybrid state $p_1$}
The hybrid state with $J=1$ and $L=J=1$ feels the potential $ \Pi_u$. It can be studied with a decoupled Schroedinger equation, with the only wave-function  $P^{0}_{1}$ giving rise to the hybrid states $(np_1)$. It is a state $J^{PC}=1^{++}$ at spin 0. Its spin-flip decay to quarkonium has one possible channel to $L'=1$ with wave-function $R_{1}$ labeled as $(np)$. The expression for the decay {width} is:
\begin{align}
\Gamma(p_1^{\text{spin}0}\to p^{\text{spin}1})=\frac{4 \alpha_s}{3 N_C} T_F \frac{\Delta E ^3}{m_Q^2} \times \left(\frac{1}{3}+1+\frac{5}{3}\right) \times \left[ \int P^0_1(r)  \ R_{1}(r) \ dr\right]^2,
    \label{eq:full_decayJ0b}
\end{align}
In the parenthesis, the $1/3$ count for the final state $J'=0$, the $1$ for the final state $J'=1$ and $5/3$ for $J'=2$.
\subsubsection{The decay of the hybrid state $(s/d)_1$}
The hybrid state with $J=1$ and $L=J\pm1=0,2$ feels the potential $ \Pi_u$ and $\Sigma_u^+$. It can be studied with a coupled Schroedinger equation, with the wave-function  $P_1^{+}/P_1^{-}$ giving rise to the hybrid states $n(s/d)_1$. It is a state $J^{PC}=1^{--}$ at spin 0. Its spin-flip decay to quarkonium has two possible channels to $L'=0$ and $L'=2$ with wave-function $R_{0}$ and $R_{2}$ labeled as $(ns)$ and $(nd)$ respectivelly. The expression for the decay {width} is:
\begin{equation}
\Gamma[(s/d)_1^{\text{spin}0}\to s^{\text{spin}1}]=\frac{4 \alpha_s}{3 N_C} T_F \frac{\Delta E ^3}{m_Q^2} \times 3 \times \left[ \int P^-_1(r)  \ R_{0}(r) \ dr\right]^2,
    \label{eq:full_decayJ1}
\end{equation}
\begin{align}
\Gamma[(s/d)_1^{\text{spin}0}\to d^{\text{spin}1}]=\frac{4 \alpha_s}{3 N_C} T_F \frac{\Delta E ^3}{m_Q^2} \times \left(\frac{3}{5}+1+\frac{7}{5}\right) \times \left[ \int P^+_1(r)  \ R_{2}(r) \ dr\right]^2.
    \label{eq:full_decayJ1b}
\end{align}
In the parenthesis, the $3/5$ count for the final state $J'=1$, the $1$ for the final state $J'=2$ and $7/5$ for $J'=3$.

\subsection{Spin 1 hybrid to spin 0 quarkonium}
We can express the wave-function for the spin 1 hybrid in terms of tensor spherical harmonics Eq. \eqref{TensorH}, and the spin 0 quarkonium state in terms of scalar spherical harmonics Eq. \eqref{S}. Then the amplitude is
\begin{align}
S^{*}H^{nk}= \int &\frac{1}{r} \sum_{\nu\mu} C(J,1,{\cal J}|{\cal M}-\nu,\nu)
C(L,1,J|{\cal M}-\nu-\mu,\mu)Y_L^{\m-\nu-\mu} 
{\chi}_\mu^n{\chi}_\nu^k P^{LJ}_{\j\m}(r) \frac{R_{L'}(r)}{r}{Y^*}_{L'}^{M'} \ r^2 dr d\Omega 
\end{align}
and the angular integral for the transition is
\begin{align}
{\cal K }^{LJ\ nk}_{ \j \m\to L'M'}\equiv\sum_{\mu\nu} C(J,1,{\cal J}| \m-\nu,\nu)C(L,1,J|\m-\nu-\mu,\mu)\times \delta_{L'}^{L}
\delta^{\m-\nu-\mu}_{M'} \widehat{\chi}_\mu^n\widehat{ \chi}_\nu^k
\end{align}
we recall that $J=|\j-1|,\j,\j+1$ and $L=|J-1|,J,J+1$ that correspond to the hybrid components of the wave-function. For each $\j$ we get up to $3^2=9$ possibilities that contribute to the amplitude. Hence, the squared amplitude may have up to $9^2=81$ terms, the non-crossed contributions shall read as
\begin{align}
[{\cal K }^{LJ}_{ \j \m\to L'}]^2\equiv {\cal K }^{LJ\ nk\,\,^\ast}_{ \j \m\to L' M'}{\cal K }^{LJ\ nk}_{ \j \m\to L' M'}=\sum_{\mu\nu } 
C^2(J,1,\j| \m -\nu,\nu)C^2(L,1,J|\m -\nu -\mu,\mu)\times \delta_{L'}^{L}\delta^{\m-\nu-\mu}_{M'}
\end{align}
{\bf Crossed terms:}
For a given $\j$ we may have some crossed terms corresponding to the components of the hybrid wave-functions that shall read as:
\begin{align}
[{\cal K }^{L_1J_1 \ nk}_{ \j \m\to L'M'}][{\cal K }^{L_2J_2\ nk}_{ \j \m\to L'M'}]=\sum_{\mu} 
&C(J_1,1,{\cal J}| \m-\nu,\nu)
C(J_2,1,{\cal J}| \m-\nu,\nu)\nonumber\\
&C(L_1,1,J_1|\m-\nu-\mu,\mu)
C(L_2,1,J_2|\m -\nu-\mu,\mu)
\times 
\delta_{L'}^{L_1}
\delta_{L'}^{L_2},
\delta^{\m-\nu-\mu}_{M'}
\end{align}
crossed terms arise from more than one contribution to the amplitude $\j J L \to L'$. This can only happen for the states $L=J\pm 1$, since they have two comments in the wave-function. However, angular momentum conservation implies $J+1=L'=J-1$ what is no accessible. 

In the {\bf spin average} squared amplitude only contribute the non-crossed terms and shall read as
\begin{align}
[{\cal K }^{LJ}_{ \j\to L'}]^2&=\frac{1}{2\j+1}\sum_{\m=-\j}^{\j}\sum_{M'=-L'}^{L'}[{\cal K }^{LJ}_{ \j \m\to L' M'}]^2 
\end{align}
Putting all together, the general expression for the transition rate is
\begin{equation}
    \Gamma(H_{\j J}^{ \text{spin}1}\to S_{L'}^{\text{spin}0})=\frac{4 \alpha_s}{3 N_C} T_F \frac{\Delta E ^3}{m_Q^2} \sum_L \ [{\cal K}^{ LJ}_{\j\to L'}]^2 \left[ \int P^{LJ}_{1\j}(r)  \ R_{L'}(r) \ dr\right]^2
    \label{eq:full_decay2b}
\end{equation}
In the following sections we apply this formula to the hybrid states $p_0$, $p_1$ and $(s/d)_1$. The results 
can be found in Table \ref{tab:decay_flipped1} and the particular values for ${\cal K}^{ LJ}_{\j\to L'}$ in Table \ref{tab:selection_rules3}. In order to estimate the reliability of our calculation  
we use in Table \ref{tab:decay_flipped1},
\begin{equation}\label{eq:<>1}
    {|\langle H_{{\cal J}J}|S_{L'}\rangle|}
    = \ \sum_L \left|\mathcal{K}^{LJ}_{\j\to L'} 
     \int P^L_J(r)  \ R_{L'}(r) \ dr\right|
\end{equation}

\begin{table}
\centering
\scalebox{0.8}{
\begin{tabular}{|c|c|c|c|c|c|} \hline
\rowcolor[gray]{0.8} 
$\j$ & $J$ & $L=L'$&$[\mathcal{K}^{LJ}_{\j\to L'}]^2$& wave function & state  \\  \hline
 0 & 0 & - & $[\mathcal{K}^{00}_{\j\to L'}]^2=$ - & $P_{0}^{00}\sim P_0^0=\nexists$ & - \\ 
 0 & 0 & 1 & $[\mathcal{K}^{+0}_{\j\to L'}]^2={0}$ & $P_{0}^{+0}\sim P_0^+$ & $p_0$ \\ \hline
 0 & 1 & 0 & $[\mathcal{K}^{-+}_{\j\to L'}]^2=1$ & $P_{0}^{-+}\sim P_1^-$ & $(s/d)_1$ \\ 
 0 & 1 & 1 & $[\mathcal{K}^{0+}_{\j\to L'}]^2=1$ & $P_{0}^{0+}\sim P_1^0$ & $p_1$ \\ 
 0 & 1 & 2 & $[\mathcal{K}^{++}_{\j\to L'}]^2=1$ & $P_{0}^{++}\sim P_1^+$ & $(s/d)_1$ \\ \hline
 1 & 0 & - & $[\mathcal{K}^{0-}_{\j\to L'}]^2=$ - & $P_{1}^{0-}\sim P_0^0=\nexists$ & -  \\ 
 1 & 0 & 1 & $[\mathcal{K}^{+-}_{\j\to L'}]^2=1$ & $P_{1}^{+-}\sim P_0^+$ & $p_0$ \\ \hline
 1 & 1 & 0 & $[\mathcal{K}^{-0}_{\j\to L'}]^2=1$ & $P_{1}^{-0}\sim P_1^-$ & $(s/d)_1$ \\ 
 1 & 1 & 1 & $[\mathcal{K}^{00}_{\j\to L'}]^2=1$ & $P_{1}^{00}\sim P_1^0$ & $p_1$ \\ 
 1 & 1 & 2 & $[\mathcal{K}^{+0}_{\j\to L'}]^2=1$ & $P_{1}^{+0}\sim P_1^+$ & $(s/d)_1$ \\ \hline
 1 & 2 & 1 & $[\mathcal{K}^{-+}_{\j\to L'}]^2=1$ & $P_{2}^{-+}\sim P_2^-$ & $(p/f)_2$ \\ 
 1 & 2 & 2 & $[\mathcal{K}^{0+}_{\j\to L'}]^2=1$ & $P_{2}^{0+}\sim P_2^0$ & $d_2$ \\ 
 1 & 2 & 3 & $[\mathcal{K}^{++}_{\j\to L'}]^2=1$ & $P_{2}^{++}\sim P_2^+$ & $(p/f)_2$ \\ \hline
\end{tabular}
}
\caption{\small \label{tab:selection_rules3} Table of selection rules for spin-flip transitions: spin 1 hybrid to spin 0 quarkonium.  {We perform the approximation $P_{\j}^{L J}\simeq P_J^{L}$} 
in the wave-functions. We mark with ``-" the cases that do not exist.}
\end{table}

\subsubsection{The decay of the hybrid state $p_0$}
The decay of the hybrid state $p_0$ arises from the combinations of quantum numbers ${\j J {L}}={001}$ and ${101}$. It can be studied with a coupled Schroedinger equation with wave-function components $P^{+0}_{0}$ and $P^{+-}_{1}$, which may be approximated to $P^{+}_0$ . Its spin-flip decay to quarkonium has $L'=1$ angular momentum conservation with wave-function $R_{1}$, labeled as $(np)$. The expression for the {decay width is}:
\begin{align}
    \Gamma(p_0^{\text{spin1}}\to p^\text{spin0})&=\frac{4 \alpha_s}{3 N_C} T_F \frac{\Delta E ^3}{m_Q^2} \times [1] \times \left[\int P^{+}_0(r)  \ R_{1}(r) \ dr\right]^2
    \label{eq:full_decayJ0c}
\end{align}

\subsubsection{The decay of the hybrid state $p_1$}
The decay of the hybrid state $p_1$ arises from the combinations of quantum numbers ${\j J L}={011}$ and ${111}$. It can be studied with a coupled Schroedinger equation with wave-function components $P^{0+}_{0}$ and $P^{00}_{1}$ approximated to $P^{0}_1$. Its spin-flip decay to quarkonium has $L'=1$ angular momentum conservation with wave-function $R_{1}$, labeled as $(np)$. The expression for the decay {width} is:
\begin{align}
    \Gamma(p_1^{\text{spin1}}\to p^\text{spin0})&=\frac{4 \alpha_s}{3 N_C} T_F \frac{\Delta E ^3}{m_Q^2} \times [1] \times \left[  \int P^{0}_1(r)  \ R_{1}(r) \ dr\right]^2
\end{align}

\subsubsection{ The decay of the hybrid state $(s/d)_1$}
The decay of the hybrid state $(s/d)_1$ arises from the combinations of quantum numbers ${\j J L}={010}$, ${012}$, ${110}$ and ${112}$. It can be studied with a coupled Schroedinger equation with wave-function components 
$P^{-+}_{0}$ and $P^{-0}_{1}$ approximated to $P^{-}_1$, plus 
$P^{++}_{0}$ and $P^{+0}_{1}$ approximated to $P^{+}_1$ . Its spin-flip decay to quarkonium has $L'=0$ or $2$ angular momentum conservation with wave-function $R_{0}$ or $R_{2}$, labeled as $(np)$ or $(nd)$. 
The expression for the decay {width} is:
\begin{align}
    \Gamma[(s/d)_1^{ \text{spin1}}\to s^{\text{spin0}}]&=\frac{4 \alpha_s}{3 N_C} T_F \frac{\Delta E ^3}{m_Q^2} \times [1] \times \left[  \int P^{-}_1(r)  \ R_{0}(r) \ dr\right]^2 \\
    \Gamma[(s/d)_1^{ \text{spin1}}\to d^{\text{spin0}}]&=\frac{4 \alpha_s}{3 N_C} T_F \frac{\Delta E ^3}{m_Q^2} \times [1] \times \left[  \int P^{+}_1(r)  \ R_{2}(r) \ dr\right]^2
\end{align}
\begin{table}[H]
\centering
\begin{minipage}{0.49 \textwidth}
\resizebox{\textwidth}{!}{
\begin{tabular}{|cc|cc|cccc|}
\hline
\rowcolor{cyan!20}
\multicolumn{8}{|c|}{Hybrid [$\mathbb S$=0] $\quad\to\quad$ Quarkonium [$\mathbb S$=1] }  \\ \hline
\rowcolor{cyan!20}
\multicolumn{2}{|c|}{Hybrid}    & \multicolumn{2}{|c|}{Quarkonium}             & \multicolumn{4}{|c|}{Decay}           \\
\rowcolor{cyan!20}
$nL_J$     & $E$   & $n'L'$ & $E'$  & $\Delta E$ & $\langle H|S \rangle \Delta E/m_Q$ & ${\langle i| r|f\rangle} \Delta E$ & $\Gamma\pm \delta \Gamma$ \\ \hline
\multicolumn{8}{|c|}{\cellcolor{lightgray}\textbf{Charm}}  \\ \hline
$2(s/d)_1$                  & 4394  & $1s$ & 3068  & 1326  & 0.56 & 0.33 & $33\pm 36$ \\
$3(s/d)_1$                  & 4678  & $1s$ & 3068  & 1610  & 0.05 & 0.57  & $0.3\pm 0.3$ \\
\hline
\multicolumn{8}{|c|}{\cellcolor{lightgray}\textbf{Bottom}}  \\ \hline
$2p_0$                      & 11261 & $1p$ & 9879  & 1382  & 0.12 & 0.08 & $1.5 \pm 1.4$  \\
$3p_0$                      & 11525 & $1p$ & 9879  & 1646  & 0.06 & 0.18&  $0.5\pm 0.4$  \\
\hline
$1(s/d)_1$                  & 10704 & $1s$ & 9551  & 1153  & 0.32 &  1.43  & $ 10\pm 7$ \\
$2(s/d)_1$                  & 10905 & $1s$ & 9551  & 1354  & 0.24 & 0.65  & $6\pm 4$ \\
$3(s/d)_1$                  & 11103 & $1s$ & 9551  & 1552  & 0.10 & 0.26  & $1\pm 0.6$ \\
\hline
\end{tabular}}
\subcaption{\label{tab:decay_flipped0} Spin-flip transitions for initial hybrid spin 0 to final quarkonium spin 1.}
\end{minipage}
\hfill
\begin{minipage}{0.49 \textwidth}
\centering
\resizebox{\textwidth}{!}{
\begin{tabular}{|cc|cc|cccc|}
\hline
\rowcolor{cyan!20}
\multicolumn{8}{|c|}{Hybrid [$\mathbb S$=1] $\quad\to\quad$ Quarkonium [$\mathbb S$=0] }  \\ \hline
\rowcolor{cyan!20}
\multicolumn{2}{|c|}{Hybrid}    & \multicolumn{2}{|c|}{Quarkonium}             & \multicolumn{4}{|c|}{Decay}           \\
\rowcolor{cyan!20}
$nL_J$     & $E$   & $n'L'$ & $E'$  & $\Delta E$  & $\langle H|S \rangle \Delta E/m_Q$ & ${\langle i|r|f \rangle} \Delta E$&$\Gamma \pm \delta \Gamma$ \\ \hline
\multicolumn{8}{|c|}{\cellcolor{lightgray}\textbf{Charm}}     \\ \hline
$2(s/d)_1$ & 4394  & $1s$ & 3068  & 1326  & 0.33 & 0.33& $11\pm 12$\\
$3(s/d)_1$ & 4678  & $1s$ & 3068  & 1610  & 0.03 &0.57 &  $0.1\pm 0.1$\\
\hline
\multicolumn{8}{|c|}{\cellcolor{lightgray}\textbf{Bottom}}     \\ \hline
$2p_0$     & 11261 & $1p$ & 9879  & 1382  & 0.07 & 0.08 &$0.5\pm 0.5$ \\
$3p_0$     & 11525 & $1p$ & 9879  & 1646  & 0.04 &0.18  &$0.2\pm 0.1$ \\
\hline
$3p_1$     & 11209 & $1p$ & 9879  & 1330  & 0.06 &  0.17&$0.4\pm 0.4$ \\
\hline
$1(s/d)_1$ & 10704 & $1s$ & 9551  & 1153  & 0.18 &   1.43 &$3.2\pm 2.5$ \\
$2(s/d)_1$ & 10905 & $1s$ & 9551  & 1354  & 0.14 & 0.65  & $2\pm 1$\\
$3(s/d)_1$ & 11103 & $1s$ & 9551  & 1552  & 0.06 & 0.26 & $0.3\pm 0.2$\\
\hline
\end{tabular}}
\subcaption{\label{tab:decay_flipped1} Spin-flip transitions for initial hybrid spin 1 to final quarkonium spin 0.}
\end{minipage}
\caption{\small \label{tab:decay_flipped}Table of spin-flip transitions for hybrid to quarkonium. Units {are} in MeV. {We only show the transitions that satisfy the constraints {(see the discussion below Eq. \eqref{eq:multp_exp:spin-flip}) and have an error $\lesssim 100\%$}.
For final quarkonium with spin 1, we present the inclusive decay rates {(the sum over possible total angular momentum $J'$ in the final state)}.
{The last column includes theoretical error associated to imprecision in $\Delta E$, higher orders in $\alpha_s$,  the multipole expansion, relativistic corrections, and the difference between weak and strong coupling regimes}, see Appendix \ref{sec:uncertainities}.}}
\end{table}

\section{Comparison with experiment}
\label{sec:comparision}
We compare our theoretical predictions with the experimental results reported by the PDG in the section ``Spectroscopy of Mesons Containing Two Heavy Quarks''~\cite{ParticleDataGroup:2024cfk}, updated through July 2025. Our analysis includes all heavy mesons listed with zero isospin, which we match to hybrid states with compatible masses and quantum numbers $J^{PC}$. Some XYZ mesons may correspond to more than one possible theoretical state. We use decay pattern compatibility to distinguish between hybrid and conventional quarkonium assignments. 
We present below our hybrid assignments. The uncertainty in this section does not include the $\delta\Delta E$ term from Eq. \eqref{eq:deltaGamma} since we  replace the mass of involved  states with the mass of the assigned PDG resonances.

\begin{table}[H]
\centering
\scalebox{0.83}{
\begin{tabular}{|c|c|c|c||c|cc|c|}
\hline
\multicolumn{4}{|c||}{\cellcolor{cyan!20}\textbf{PDG}} & \multicolumn{4}{|c|}{\cellcolor{cyan!20}\textbf{ Our estimate}} \\ \hline
\hline
\rowcolor{cyan!20}
Name & Mass & $J^{PC}$ & $\Gamma$ & Hybrid  & \multicolumn{2}{|c|}{Quarkonium decay channel}&$\Gamma_{q\bar{q}g\to q\bar{q}}$ \\ 
\hline
\multicolumn{8}{|c|}{\cellcolor{lightgray}\textbf{Charm}} \\ \hline
$X(3940)$ &  ${3942\pm 9}$ & $?^{??}$ & $43\pm 20$ & $1(s/d)_1[\mathbb{S}=1]$  
&\makecell[tc]{$1s[\mathbb{S}=0]$\\$1p[\mathbb{S}=1]$} 
&\makecell[tc]{\\ }
&\makecell[tc]{NR\\NR} \\  \hline
$\psi(4040)$ &  ${4040\pm 4}$ & $1^{--}$ & $84\pm 12$ & $1(s/d)_1[\mathbb{S}=0]$  &\makecell[tc]{$1p[\mathbb{S}=0]$\\ $1s[\mathbb{S}=1]$} &\makecell[tc]{  \\ }&\makecell[tc]{NR\\NR} \\ \hline
$\chi_{c1}(4140)$ &  ${4146\pm 3}$ & $1^{++}$ & $19\pm 7$ & $1p_1[\mathbb{S}=0]$  
&\makecell[tc]{$1s[\mathbb{S}=0]$\\$1p[\mathbb{S}=1]$} 
&\makecell[tc]{\\}
&\makecell[tc]{NR\\NR} \\ \hline
$X(4160)$ &  ${4153\pm 20}$ & $?^{??}$ & $136^{+ 60}_{-35}$ & $1p_1[\mathbb{S}=1]$  
& \makecell[tc]{$1p[\mathbb{S}=0]$\\$1s[\mathbb{S}=1]$}
& 
&\makecell[tc]{NR\\LC} \\ \hline
$X(4350)$ &  ${4351\pm 5}$ & $?^{?+}$ & $13^{+18}_{-10}$ & $2(s/d)_1[\mathbb{S}=1]$  
&\makecell[tc]{$1s[\mathbb{S}=0]$\\$1p[\mathbb{S}=1]$}
&\makecell[tc]{$\eta_c(1S) \ [2984]$\\ }
&\makecell[tc]{$12\pm 12$\\LC} \\ \hline
$\psi(4360)$ &  ${4374\pm 7}$ & $1^{--}$ & $120\pm12$ & $2(s/d)_1[\mathbb{S}=0]$  
&\makecell[tc]{$1p[\mathbb{S}=0]$\\$1s[\mathbb{S}=1]$} 
&\makecell[tc]{\\$J/\psi(1S) \ [3097]$}
&\makecell[tc]{NR\\$30\pm 32$ } \\ \hline
$\psi(4415)$ &  ${4415\pm 5}$ & $1^{--}$ & $110\pm13$ & $2(s/d)_1[\mathbb{S}=0]$  &\makecell[tc]{$1p[\mathbb{S}=0]$\\$1s[\mathbb{S}=1]$} &\makecell[tc]{\\$J/\psi(1S) \ [3097]$}&\makecell[tc]{ NR\\$32\pm33$} \\ \hline
{$\chi_{c0}(4500)$} &  ${4474\pm 4}$ & $0^{++}$ & $77\pm 10$ & $1p_0[\mathbb{S}=0]$  
& \makecell[tc]{$ 1s[\mathbb{S}=0]$\\ $ 1p[\mathbb{S}=1]$} 
&\makecell[tc]{\\}
&\makecell[tc]{NR\\LC}\\ \hline
$X(4630)$ &  ${4626^{+24}_{-110}}$ & $?^{?+}$ & $170^{+140}_{-80}$ & $3(s/d)_1[\mathbb{S}=1]$  
&\makecell[tc]{$1s[\mathbb{S}=0]$\\$1p[\mathbb{S}=1]$ }
&\makecell[tc]{$\eta_c(1S)\ [2984]$\\}
&\makecell[tc]{$0.12\pm 0.1$\\LC} \\ \hline  
$\psi(4660)$ &  ${4623\pm 10}$ & $1^{--}$ & $55\pm 9$ & $3(s/d)_1[\mathbb{S}=0]$ 
&\makecell[tc]{$1p[\mathbb{S}=0]$\\$1s[\mathbb{S}=1]$} 
&\makecell[tc]{\\$J/\psi(1S) \ [3097]$} 
&\makecell[tc]{NR\\$0.3\pm 0.27$} \\  \hline
\multicolumn{8}{|c|}{\cellcolor{lightgray}\textbf{Bottom}} \\ \hline
$\Upsilon(10753)$ &  ${10753\pm 6}$& $1^{--}$ & $36\pm 15$ & $1(s/d)_1[\mathbb{S}=0]$  
& \makecell[tc]{$1p[\mathbb{S}=0]$\\$1s[\mathbb{S}=1]$}
&\makecell[tc]{\\$\Upsilon(1S) \ [9460]$}
&\makecell[tc]{NR\\$  13\pm9$} \\ \hline
$\Upsilon(10860)$ &  ${10885\pm 2}$ & $1^{--}$ & $37\pm 4$ & $2(s/d)_1[\mathbb{S}=0]$  
& \makecell[tc]{$1p[\mathbb{S}=0]$\\$1s[\mathbb{S}=1]$}
& \makecell[tc]{$h_b(1P) \ [9899]$\\$\Upsilon(1S) \ [9460]$\\}
&\makecell[tc]{LC\\$7\pm4$\\} \\ \hline
$\Upsilon(11020)$ &  ${11000\pm 4}$& $1^{--}$ & $24\pm 7$ & $3(s/d)_1[\mathbb{S}=0]$  
&\makecell[tc]{$2p[\mathbb{S}=0]$\\$1s[\mathbb{S}=1]$}
& \makecell[tc]{$h_b(2P) \ [10260]$\\$\Upsilon(1S) \ [9460]$}
&\makecell[tc]{LC\\$1\pm0.6$} \\ \hline
\end{tabular}
}
    \caption{\label{tab:XYZ}\small {Table} for XYZ mesons, as presented by the PDG, and the corresponding states {from our calculation}. All units are in MeV. Hybrids are labeled by their quantum numbers $nL_J[\mathbb{S}]$, while quarkonium states are labeled as $nL[\mathbb{S}]$. The values of $\Gamma$, along with their associated uncertainties, are refined by replacing {our calculated masses in $\Delta E$ } with the corresponding experimental value. Uncertainties include higher-order corrections in $\alpha_s$ and in the multipole expansion, relativistic corrections, {and the difference between weak and strong coupling regimes}.
    We use the notation Not Reliable (NR) {if $\Delta E\lesssim 800$ MeV or if $\langle H|r|S\rangle\Delta E\gtrsim 1$ ($\langle H|S \rangle\Delta E/m_Q\gtrsim 1$) for spin-conserved (spin-flip) transitions.   
    Large Corrections (LC) means $\delta  \Gamma \gtrsim 100\%$, {which is discussed in Appendix \ref{sec:uncertainities}. 
    For details of each transition, we refer the reader to the table \texttt{decay\_tables\_full.pdf} in the supplementary material Ref.~\cite{oncala_mesa_2025_hybrid}.}}
}
\label{tab:hybrids} 
\end{table}

\begin{itemize}
    \item $X(3940)$ ($J^{PC} = ?^{??}$, $\Gamma = 43 \pm 20~\text{MeV}$). 
    It can be identified with our hybrid $1(s/d)_1$ ($4028$ MeV) in any spin configuration. 
    We find that neither the spin-conserved nor the spin-flip decays can be calculated reliably. An assignment as $2p$  ($3958$ MeV) charmonium remains possible.

    \item $\psi(4040)$ ($J^{PC} = 1^{--}$, $\Gamma = 84 \pm 12~\text{MeV}$). 
    It can be identified as our hybrid $1(s/d)_1$ ($4028$ MeV) with spin~0. 
    We find that neither the spin-conserved nor the spin-flip decays can be calculated reliably. If this identification is made, then $X(3940)$ can be one of the spin-1 partners, most likely the $0^{-+}$, according to hyperfine splitting pattern found in \cite{Brambilla:2018pyn,Brambilla:2019jfi,Soto:2023lbh}. 

    \item $\chi_{c1}(4140)$ ($J^{PC} = 1^{++}$, $\Gamma = 19 \pm 7~\text{MeV}$). 
    It can be identified as our hybrid $1p_1$ ($4171$ MeV) with spin~0\footnote{In ref. \cite{Oncala:2017hop} this identification was ruled out because of a selection rule which was due to an oversimplification in the derivation of the spin-conserved transitions\cite{Brambilla:2022hhi}.\label{over}}. 
    We find that neither the spin-conserved nor the spin-flip decays can be calculated reliably. 
    
    \item  $X(4160)$ ($J^{PC} = ?^{??}$, $\Gamma = 136^{+90}_{-35}$ MeV). 
    It may be identified as our hybrid  $1p_1$  ($4171$ MeV) at any spin configuration. If $\chi_{c1}(4140)$ is assigned to the spin-0 $1p_1$ hybrid, then $X(4160)$ can be one of the spin-1 partners, most likely the $1^{-+}$, according to the hyperfine splitting pattern found in \cite{Brambilla:2018pyn,Brambilla:2019jfi,Soto:2023lbh}.
    We find that neither the spin-conserved nor the spin-flip decays can be reliably calculated.
    An assignment as $3s$ ($4149$ MeV) or $2d$ ($4209$ MeV) charmonium, or $1(p/f)_2$ ($4245$ MeV) hybrid remains possible.

    \item   $X(4350)$ ($J^{PC} = ?^{?+}$, $\Gamma =13^{+18}_{-10}~\text{MeV}$).
    It can be identified as the hybrid $2(s/d)_1$ ($4394$ MeV) at spin-1. 
    {We find  a} spin-flip decay  into charmonium $\eta_c(1S)$, with a  width of $12\pm {12}$ MeV. {Even though this value is consistent with the total decay width, it leaves little room for spin-conserved transitions and for decays to $D$-meson pairs}. 
    An assignment as $3p$ ($4388$ MeV) spin-1 charmonium {is favored, and an assignment} as $1p_0$ ($4455$ MeV) spin-0 hybrid
    remains possible.

    \item  $\psi(4360)$ and $\psi(4415)$, ($J^{PC} = 1^{--}$, $\Gamma =120 \pm 12~\text{MeV}$ and $\Gamma=110 \pm 13~\text{MeV}$, respectively). 
    They both can be identified as our hybrid $2(s/d)_1$ ($4394 $ MeV) at spin-0. 
    We find{ a} spin-flip decay into charmonium $J/\psi(1S)$, with a { width} of $\Gamma=30\pm {32}$ MeV and $\Gamma=32\pm {33}$ MeV, respectively, { which is consistent} with { the observed total decay width in both cases. }

    \item  $\chi_{c0}(4500)$ ($J^{PC} = 0^{++}$, $\Gamma=77 \pm 10~\text{MeV}$). 
    It can be identified with our $1p_0$ hybrid  ($4445$ MeV) at spin-0. 
    We find that neither the spin-conserved nor the spin-flip decays can be reliably calculated. 
    An assignment to the nearby  $3p$ quarkonium ($4388$ MeV) is less favored.

    \item $X(4630)$ ($J^{PC} = ?^{?+}$, $174 \pm 78~\text{MeV}$). 
    It can be identified as our hybrid $3(s/d)_1$ ($4678$ MeV) at spin-1. 
    {We find  a} spin-flip decay into charmonium $\eta_c(1S)$, with a  width of $\Gamma\sim 0.12\pm{0.1}$ MeV,  which is consistent with the observed total decay width.
    An assignment as $4s$ ($4562$ MeV) or  $3d$ ($4608$ MeV) spin-0  charmonium remains possible. 
    
    \item $\psi(4660)$ ($J^{PC} = 1^{--}$, $\Gamma =55 \pm 9~\text{MeV}$).
    It can be identified as our hybrid $3(s/d)_1$ ($4678$ MeV) at spin-0. 
    {We find a} spin-flip transition channel into charmonium $J/\psi(1S)$ with a width of $\Gamma\sim 0.3\pm0.27$ MeV, { which is consistent with the observed total decay width}.
    An assignment as {$4s$ ($4562$ MeV) or} $3d$ ($4608$ MeV) spin-1 charmonium remains possible.

\newcommand{\hybridnote}{
  \footnote{For the three bottomonium hybrids $1(s/d)_1$, $2(s/d)_1$ and $3(s/d)_1$, our mass predictions do not match well with the experimental values. This mismatch might be due to strong mixing effects, as the $(s/d)_1$ hybrid states lie close in energy to the nearby $s$ and $d$ bottomonium levels \cite{Oncala:2017hop}. Such mixing could shift the predicted mass positions and alter decay widths accordingly.
  }
}
    
    \item $\Upsilon(10753)$ ($J^{PC} = 1^{--}$, $\Gamma =36 \pm 15~\text{MeV}$). 
    It {may} be identified\hybridnote $\,$ 
     as our hybrid state $1(s/d)_1$ ($10704$ MeV) at spin-0 ($49$ MeV $\sim 1 \sigma$ away). 
    We find a spin-flip decay  into bottomonium $\Upsilon(1S)$, with a  width of $\Gamma\sim 13\pm 9$ MeV, which is consistent with the observed total decay width.
    An assignment as bottomonium $3d$ ($10689$ MeV) spin-1 {($64$ MeV $\sim$ {$1\sigma$ away) may also be} possible. 
    
    \item $\Upsilon(10860)$ (with $J^{PC} = 1^{--}$, $\Gamma =37 \pm 4~\text{MeV}$). 
    It {may} be identified\footnotemark[\value{footnote}]
   as our hybrid  $2(s/d)_1$ ($10905$ MeV) at spin-0.
    { We find  a} spin-flip decay into bottomonium $\Upsilon(1S)$, with a width of $\Gamma= 7\pm4$ MeV. {Since the total decay width to non-$B\bar B$ states is $\Gamma_{{\rm non-}B\bar B}=6.5\pm 1$ GeV, our numbers disfavor the assignment as a pure hybrid state}. 
    An assignment as  bottomonium $5s$ ($10901$ MeV) is favored.

    \item $\Upsilon(11020)$ ($J^{PC} = 1^{--}$, $\Gamma=24 \pm 5~\text{MeV}$).
    It {may} be identified\footnotemark[\value{footnote}] 
     as our hybrid $3(s/d)_1$ ($11103$ MeV) at spin-0 ($103$ MeV $\sim 3 \sigma$ away}). {We find  a} spin-flip decay into bottomonium $\Upsilon(1S)$, with a  width of $\Gamma= 1\pm0.6$ MeV. 
    An assignment as bottomonium $4d$ ($10939$ MeV) ($61$ MeV $\sim$$1 \sigma$ away) is favored. 
\end{itemize}

\section{Summary and Conclusions}
\label{sec:conclusion}

 Within the Born-Oppenheimer Effective Field Theory (BOEFT) framework we update and extend the results of Ref. \cite{Oncala:2017hop}. In particular we revise the calculation for spin-conserved transitions and address the spin-flip ones \cite{TarrusCastella:2021pld,Brambilla:2022hhi}. Our analysis extends previous works \cite{Juge:1999ie,Braaten:2014qka,Berwein:2015vca,Oncala:2017hop,Brambilla:2022hhi} in several important ways:

\begin{itemize}
\item \textbf{Updated hybrid potentials.} We incorporate a recent evaluation of the hybrid static potentials \cite{Alasiri:2024nue}, which uses the latest lattice QCD data \cite{Capitani:2018rox,Schlosser:2021wnr}. This 
improves the accuracy of the energy spectrum and wavefunctions used in our decay width calculations. The differences with respect to the energy spectrum reported in \cite{Oncala:2017hop} fall within the expected theoretical uncertainties.

\item \textbf{Spin-flip transitions.} We treat spin-flip transitions with greater care, particularly those originating from spin-0 hybrids and leading to spin-1 quarkonium final states, which form multiplets with different total angular momenta $J$. We calculate the decay width for every $J$, which allows us to make precise comparisons with PDG data.  In ref. \cite{Brambilla:2022hhi} only the sum of decay widths for all $J$ is displayed.

\item \textbf{Inclusion of the $d$-wave decay channel.} We include the $d$-wave {final states} in the decay channel  both in spin-conserved and spin-flip transitions, which have not been considered before. 
This channel allows transitions from the hybrid initial state $p_J$ through spin-conserving processes and from $(s/d)_1$ through spin-flip ones. Nevertheless, {the figures obtained for these transitions turn out to be unreliable (either because the transitions do not fulfill the constraints or because the figures are produced with large errors, see the supplemental material Ref.~\cite{oncala_mesa_2025_hybrid}).} 

\item \textbf{Comprehensive error analysis for the decay widths.} We make a systematic analysis of errors, including theoretical errors associated to higher orders in the strong coupling, imprecision in the mass spectrum, the multipole expansion, the relativistic corrections, and the difference between weak and strong coupling regimes. The last error source turns out to be very important. In ref. \cite{Brambilla:2022hhi}, only the first three error sources are considered. {As a consequence, a number of decay widths figures that are used to rule out possible hybrid candidates in that reference become unreliable once the remaining error sources are taken into account, in particular the last one.  In ref. \cite{Oncala:2017hop}, the relativistic corrections are not discussed either.  
Even though $\Delta E$ are comparable to $m_c$ in quite a few cases, the relativistic corrections do not have a major impact in the final error.
However, due to numerical factors, in most cases the error from the relativistic corrections turns out to be more important than the one from the multipole expansion, which is parametrically larger. Hence, it is important to take it into account.}
We  carefully discuss here for which decay channels our calculation produces reliable results.
\end{itemize}

We also recalculate the spin-conserved transitions, which are now in agreement with ref. \cite{Brambilla:2022hhi}.
These improvements lead to more accurate assignments of XYZ resonances to quarkonium or hybrid states, and help resolve some ambiguities in the state identification. We confirm the assignments of ref. \cite{Oncala:2017hop} with the following exceptions: $\chi_{c1}(4140)$, which was left with no assignment is now identified with the spin zero $1p_1$ hybrid{\color{blue}\footref{over}}
; $\psi($4040$)$ and {$\psi(4415)$}
, which were assumed to be quarkonium states, are now identified with the spin zero $1(s/d)_1$\footnote{This state was identified with $Y(4008)$ in ref. \cite{Oncala:2017hop}. $Y(4008)$ was reported by Belle in refs. \cite{Belle:2007dxy,Belle:2013yex}, but it has not been confirmed by other experiments (see, for instance,\cite{BESIII:2016bnd,BaBar:2012vyb}) and it is not listed in the PDG. } and $2(s/d)_1$\footnote{It was assigned to $4s$ in ref. \cite{Oncala:2017hop}.} hybrids; $X(3940)$, which was assigned to $2p$ state, now it can also be a $1(s/d)_1$ hybrid. 

{Let us mention that
the spectrum presented here describes the higher bottomonium resonances less accurately than the one in ref. \cite{Oncala:2017hop}.
This may be due to the fact that the potential we use here is obtained entirely from fits to quenched lattice data \cite{Alasiri:2024nue}. This delivers a string tension $\sigma\sim 0.22$ GeV$^2$, whereas in ref. \cite{Oncala:2017hop} a value $\sigma\sim 0.18$ GeV$^2$ was used, which is closer to the values obtained by lattice calculations with dynamical fermions \cite{Bulava:2024jpj} (see \cite{Soto:2025spl} for a discussion on the light-quark mass dependence of the string tension). In addition, the proximity of several quarkonium $s$ and $d$ levels to hybrid $(s/d)_1$ levels may make mixing important \cite{Oncala:2017hop}. We expect that the mixing issue will be resolved once the recent lattice data on the mixing potentials of ref. \cite{Schlosser:2025tca} is incorporated in the approach.}

The usefulness of the semi-inclusive decay widths to quarkonium in order to identify hybrid states {among the XYZ resonances} is limited. This is due to the fact that only a few transitions can be reliably calculated, and they not always correspond to XYZ candidates. Nevertheless, we find, for instance, that the calculated {spin-flip decay width of $\Upsilon(10860)$ is comparable the branching fraction to non-$B\bar B$ states reported in the PDG. Since there should also be room for the decay width of the spin-conserved transitions, its interpretation as a purely hybrid state is disfavored. $X(4350)$ is a similar case: the spin-flip decay width is similar to the total decay width.} 
However, most of the calculated semi-inclusive decay widths that correspond to XYZ states are compatible with the total decay width reported in the PDG. In this case, we provide predictions for particular channels. The remaining semi-inclusive decay widths that we have calculated stand as predictions for these hybrid states, which may be eventually tested if they are identified with new XYZ resonances in the future.

\vspace{0.5em}

In summary, we keep finding an overall good agreement with the experimental data reported by the PDG~\cite{ParticleDataGroup:2024cfk}. We can assign either hybrid or conventional quarkonium interpretations to nearly all heavy isospin-zero states in the XYZ meson spectrum. In addition, we have been able to reliably estimate a number of semi-inclusive decay widths  for both spin-conserved and spin-flip hybrid to quarkonium transitions by carrying out an exhaustive error analysis.\\

{\bf Acknowledgements}
\\

JS acknowledges financial support from Grant No. 2021-SGR-249 from the Generalitat de Catalunya and from projects No. PID2022-136224NB-C21, PID2022-139427NB-I00 and No. CEX2024-001451-M from Ministerio de Ciencia, Innovaci\'on y Universidades (MICIU/AEI/10.13039/501100011033).

\clearpage

\appendix

\section{Uncertainties \label{sec:uncertainities}}

\subsection{Uncertainty in the mass spectrum}
\label{sec:uncertainity_mass}
We have used the leading-order potential for both quarkonium and hybrids; the potentials we neglect start at order $1/m_Q$. For hybrids, the error we assign to this calculation is $\Lambda_{\text{QCD}}^2/m_Q$, since $\Lambda_{\text{QCD}}$ is the next relevant scale. For quarkonium, this is not always the case, as the typical momenta can be larger than $\Lambda_{\text{QCD}}$. A detailed error analysis is carried out in Appendix A of~\cite{Oncala:2017hop}. For simplicity, we stick to $\Lambda_{\text{QCD}}^2/m_Q$ to estimate the uncertainty for quarkonium as well. Taking $\Lambda_{\text{QCD}} \sim 400~\text{MeV}$, we obtain a precision of about $110~\text{MeV}$ for charmonium and $33~\text{MeV}$ for bottomonium. \\

$\Delta E$ is the energy gap between the initial hybrid and the final quarkonium states in the transition. In most of our transitions $\Delta E$ typically lies in the range of $1$-$2~\text{GeV}$, and is never smaller than $800~\text{MeV}$. To estimate uncertainties, we vary $\Delta E$ by $\delta\Delta E=\sqrt{2} \times 110~\text{MeV}$ for charmonium and $\delta\Delta E= \sqrt{2} \times 33~\text{MeV}$ for bottomonium.

\subsubsection{Quarkonium}
We compare the quarkonium spectrum from our calculation with the particles listed by the PDG~\cite{ParticleDataGroup:2024cfk}. Since we work at leading-order in the $1/m_Q$ expansion for the potentials, the states in our calculation are spin-degenerate, so they may correspond to various $J^{PC}$ configurations for ${\mathbb S} = 0$ or ${\mathbb S} = 1$, as shown in Table~\ref{fig:hybrid_spectrum}. \\

We observe that most of the computed states do not differ from the experimental candidates by more than the expected precision, see Table \ref{tab:charmonium0}. For charmonium, we obtain a mean absolute error (m.a.e) of $57~\text{MeV}$, which is well below the expected theoretical uncertainty of $110~\text{MeV}$. In the bottomonium sector, we find a m.a.e. of $34~\text{MeV}$, slightly above the expected uncertainty of $33~\text{MeV}$. However, it is important to note that the bottomonium ground state ($1s$) is expected to deviate more significantly due to its large mass and stronger binding effects\footnote{ $\Upsilon (1S)$ and $\eta_b(1S)$ are better understood in the weak coupling regime of pNRQCD, see for instance \cite{Kniehl:2003ap,Peset:2018ria}.}. If we exclude the $(1s)$ states from the statistics, the resulting subset yields a reduced m.a.e. of $26~\text{MeV}$, which lies below the expected precision.

\subsubsection{Hybrids} 
The static potentials may be fitted to lattice data with various approximations for the long- and short-distance behavior. Variations in the assumed potential lead to differences in the predicted spectrum. In Table \ref{tab:hybrids}, we summarize the hybrid spectrum for different fits and associate each state with its most likely experimental candidate.\\

For hybrid charmonium, we obtain a m.a.e of 35~MeV, which is well below the expected theoretical uncertainty of 110~MeV. In the bottomonium sector, however, we find fewer possible identifications, and some are further from the predicted energy levels. In this case, the m.a.e. is 35~MeV, slightly above the expected uncertainty of 33~MeV. It is important to note that these statistical estimates are based only on the identified hybrid candidates (10 for charmonium and only 2 for bottomonium) so the statistical significance of the bottomonium result is limited.
\begin{SCtable}[1]
  \resizebox{0.75\textwidth}{!}{%
      \begin{subtable}[t]{0.55\linewidth}
        \begin{tabular}{|c|c||c|c|c|}
          \hline
          \rowcolor{cyan!20}
          \multicolumn{2}{|c||}{\textbf{Our prediction}} & \multicolumn{3}{|c|}{\textbf{PDG}}\\ 
          \hline\hline
          \rowcolor{cyan!20}
          $nL$ & $M_{c\bar{c}}\pm110$MeV & Mass & PDG Name & $J^{PC}$\\ \hline 
          $1s$ & 3068 & 2984 & $\eta_c(1S)$ & 0$^{-+}$ \\
               &      & 3097 & $J/\psi(1S)$ & 1$^{--}$ \\ 
          $2s$ & 3674 & 3638 & $\eta_c(2S)$ & 0$^{-+}$ \\  
               &      & 3686 & $\psi(2S)$ & 1$^{--}$ \\           
          $3s$ & 4149 & $4153^{+23}_{-21}$ & $\color{orange}X(4160)$ & ?$^{??}$ \\
               &      & 4191$\pm5$ & $\psi(4160)$ & 1$^{--}$ \\
          $4s$ & 4562 & $4626^{+24}_{-110}$ & \color{orange}$X(4630)$ & ?$^{?+}$ \\  
               &      & $\color{gray}4415\pm5$ & \color{orange}$\psi(4415)$ & 1$^{--}$ \\ \hline
          $1p$ & 3457 & 3525 & $h_c(1P)$ & 1$^{+-}$ \\
               &      & 3415 & $\chi_{c0}(1P)$ & 0$^{++}$ \\
               &      & 3511 & $\chi_{c1}(1P)$ & 1$^{++}$ \\
               &      & 3556 & $\chi_{c2}(1P)$ & 2$^{++}$ \\
          $2p$ & 3958 & $3942\pm9$ & \color{orange}$X(3940)$ & ?$^{??}$ \\ 
               &      & $3862^{+50}_{-35}$ & $\chi_{c0}(3860)$ & 0$^{++}$ \\ 
               &      & $3922\pm2$ & $\chi_{c0}(3915)$ & 0$^{++}$ \\ 
               &      & 3872 & $\chi_{c1}(3872)$ & 1$^{++}$ \\
               &      & $3922\pm1$ & $\chi_{c2}(3930)$ & 2$^{++}$ \\ 
          $3p$ & 4388 & $4474\pm4$ & \color{orange}$\chi_{c0}(4500)$ & 0$^{++}$ \\ 
               &      & $4286\pm9$ & $\chi_{c1}(4274)$ & 1$^{++}$ \\ 
               &      & $4351\pm5$ & \color{orange}$X(4350)$ & ?$^{?+}$ \\ 
          $4p$ & 4774 & $4694^{+16}_{-5}$ & $\chi_{c0}(4700)$ & 0$^{++}$ \\ 
               &      & $4684^{+15}_{-17}$ & $\chi_{c1}(4685)$ & 1$^{++}$ \\ \hline
          $1d$ & 3762 & 3774 & $\psi(3770)$ & 1$^{--}$ \\
               &      & 3823 & $\psi_2(3823)$ & 2$^{--}$ \\
               &      & 3842 & $\psi_3(3842)$ & 3$^{--}$ \\
          $2d$ & 4209 & $4222\pm2$ & $\psi(4230)$ & 1$^{--}$ \\
          $3d$ & 4608 & $4623\pm10$ & \color{orange}$\psi(4660)$ & 1$^{--}$ \\ 
          \hline
        \end{tabular}
        \subcaption{\large Charmonium spectrum and candidates.}
      \end{subtable}
      \hfill
      \begin{subtable}[t]{0.52\linewidth}
        \begin{tabular}{|c|c||c|c|c|}
          \hline
          \rowcolor{cyan!20}
          \multicolumn{2}{|c||}{\textbf{Our prediction}} & \multicolumn{3}{|c|}{\textbf{PDG}}\\ 
          \hline\hline
          \rowcolor{cyan!20}
          $nL$ & $M_{b\bar{b}}\pm33$MeV & Mass & PDG Name & $J^{PC}$\\ \hline 
          $1s$ & 9551 & $\it \color{gray}9399\pm2$ & $\eta_b(1S)$ & 0$^{-+}$ \\
               &      & $\it \color{gray}9460$ & $\Upsilon(1S)$ & 1$^{--}$ \\
          $2s$ & 10017 & 9999$\pm$4 & $\eta_b(2S)$ & 0$^{-+}$ \\
               &       & 10023 & $\Upsilon(2S)$ & 1$^{--}$ \\ 
          $3s$ & 10355 & 10330 & $\eta_b(3S)$ & 0$^{-+}$ \\  
               &       & 10355 & $\Upsilon(3S)$ & 1$^{--}$ \\ 
          $4s$ & 10643 & $\color{gray}10579\pm1$ & $\Upsilon(4S)$ & 1$^{--}$ \\  
          $5s$ & 10901 & $10885\pm$2 & $\color{orange} \Upsilon(10860)$ & 1$^{--}$ \\ \hline
          $1p$ & 9879 & 9899 & $h_b(1P)$ & 1$^{+-}$ \\ 
               &      & 9859 & $\chi_{b0}(1P)$ & 0$^{++}$ \\ 
               &      & 9893 & $\chi_{b1}(1P)$ & 1$^{++}$ \\ 
               &      & 9912 & $\chi_{b2}(1P)$ & 2$^{++}$ \\ 
          $2p$ & 10234 & 10260$\pm$1 & $h_b(2P)$ & 1$^{+-}$ \\ 
               &       & 10233 & $\chi_{b0}(2P)$ & 0$^{++}$ \\ 
               &       & 10256 & $\chi_{b1}(2P)$ & 1$^{++}$ \\ 
               &       & $\color{gray}10269$ & $\chi_{b2}(2P)$ & 2$^{++}$ \\ 
          $3p$ & 10532 & 10530 & $\chi_{b0}(3P)$ & 0$^{++}$ \\ 
               &       & 10513 & $\chi_{b1}(3P)$ & 1$^{++}$ \\ 
               &       & 10524 & $\chi_{b2}(3P)$ & 2$^{++}$ \\ \hline
          $1d$ & 10106 & $\color{gray}10164\pm1$ & $\Upsilon_2(1D)$ & 2$^{--}$ \\ 
          $3d$ & 10689 & $\color{gray}10753\pm6$ &  $\color{orange} \Upsilon(10753)$ & 1$^{--}$ \\ 
          $4d$ & 10939 & $\color{gray}11000\pm4$ & $\Upsilon(11020)$ & 1$^{--}$ \\ 
          \hline
        \end{tabular}
        \subcaption{\large Bottomonium spectrum and candidates.}
      \end{subtable}
  }
  \caption{\footnotesize Comparative table for the quarkonium spectrum (in~MeV) showing our predicted states and the experimental candidates listed by the PDG~\cite{ParticleDataGroup:2024cfk} 
  with compatible $J^{PC}$ and close mass. We only include uncertainties in PDG masses when they are bigger than 1 MeV. We consider an expected precision ($1\sigma$) of about $110\,\text{MeV}$ for charmonium and $33\,\text{MeV}$ for bottomonium.  We use the {following} color code: orange for  PDG states that could correspond to several of our candidates, either quarkonia or hybrid; gray for PDG states that are more than $1\sigma$ away from our mass prediction; {gray italics} the bottomonium ground states ($1s$), which are better understood in the weak coupling regime.}
  \label{tab:charmonium0}
\end{SCtable}
\begin{SCtable}[1.2] 
  \resizebox{0.55\textwidth}{!}{  
\begin{tabular}{|c|c|c|c||c|c|c|}
\hline
\rowcolor{cyan!20}
\multicolumn{3}{|c|}{\textbf{Our prediction}} && \multicolumn{3}{|c|}{\textbf{PDG}}\\ 
\hline
\hline
\rowcolor{cyan!20}
$nL_J$     &  $M_{q\bar{q}g}$    & $M_{q\bar{q}g}$\cite{Oncala:2017hop}      & $M_{q\bar{q}g}$\cite{Brambilla:2022hhi} & Mass  &  PDG Name & $J^{PC}$ \\  \hline
\rowcolor{lightgray}
\multicolumn{4}{|c}{$\pm110$MeV}&\multicolumn{3}{c|}{\textbf{Charm}}     \\ \hline
$1p_0$     &  4455   & 4486  & 4590   &$4474\pm4$&$\color{orange}\chi_{c0}(4500)$&$0^{++}$      \\ 
$2p_0$     &  4917   & 4920 & 5054   &&&       \\ 
$3p_0$     &  5315   & 5299& 5473    &&&      \\ \hline 
$1(s/d)_1$ &  4028   &4011  & 4155   &$4040\pm4$&$\psi(4040)$& 1$^{--}$      \\ 
           &         &     &       &$3942\pm9$&$\color{orange} X(3940)$&  ?$^{??}$   \\
$2(s/d)_1$ &  4394   &4355 & 4507    &$4374\pm7$&$\psi(4360)$&  1$^{--}$   \\ 
 &         &     &       &$4415\pm5$&$\color{orange}\psi(4415)$&  1$^{--}$   \\
            &         &     &       &$4351\pm5$&$\color{orange}X(4350)$&  ?$^{?+}$   \\
$3(s/d)_1$ &  4678   & 4692& 4812   &$4623\pm10$&{\color{orange}$\psi(4660)$}& 1$^{--}$     \\
 &     & &    &$4626^{+24}_{-110}$&$\color{orange}X(4630)$& ?$^{?+}$     \\\hline  
$1p_1$     &  4171   & 4145  & 4286  &$4146\pm3$&$\chi_{c1}(4140)$&1$^{++}$       \\ 
           &         &     &       & $4153^{+23}_{-21}$&$\color{orange}X(4160)$&  ?$^{??}$   \\
$2p_1$     &  4556   & 4511 & 4667   &&&       \\ 
$3p_1$     &  4912   & 4863 & 5035  &&&      \\ \hline 
\rowcolor{lightgray}
\multicolumn{4}{|c}{$\pm33$MeV}&\multicolumn{3}{c|}{\textbf{Bottom}}     \\ \hline
$1p_0$     &  10977   & 11011  & 11065   &&&    \\ 
$2p_0$     &  11261   & 11299 & 11352    &&&   \\
$3p_0$     &  11525   & 11551& 11616    &&&    \\ \hline 
$1(s/d)_1$ &  10704   & 10690  & 10786  &$\color{gray}10753\pm6$&$\color{orange}\Upsilon(10753)$&  1$^{--}$     \\ 
$2(s/d)_1$ &  10905   & 10885 & 10976   &$10885\pm2$&$\color{orange}\Upsilon(10860)$&  1$^{--}$    \\ 
$3(s/d)_1$ &  11103   & 11084& 11172    &&& \\ \hline
$1p_1$     &  10772   &10761 & 10846    &&&    \\ 
$2p_1$     &  10995   & 10970 & 11060   &&&      \\ 
$3p_1$     &  11209   & 11175& 11270    &&&   \\ \hline 
\end{tabular}
}
    \caption{\footnotesize Comparative table for the hybrid spectrum. For each state $nL_J$ we show the predicted energy (in~MeV) for different potentials: those used in the present work, showed in Appendix \ref{sec:stat_pot}, the one used in our previous {paper}~\cite{Oncala:2017hop} and the one used in ref. \cite{Brambilla:2022hhi}.  {The $(s/d)_1$, $p_1$, $p_0$, $(p/f)_2$ and $d_2$ states are named $H_1$, $H_2$, $H_3$, $H_4$ and $H_5$ respectively in \cite{Berwein:2015vca, Brambilla:2022hhi}.} The last columns show our best PDG candidate \cite{ParticleDataGroup:2024cfk}. The expected precision of predictions is about $110\text{ MeV}$ for charmonium and $33\text{ MeV}$ for bottomonium. In orange:  PDG states that could correspond to several of our candidates, either quarkonia or hybrid. In gray: PDG states that are more than $1\sigma$ away from our mass prediction.
\label{tab:hybrids0} }
  \end{SCtable}

\clearpage
\subsection{Uncertainty in  $\Gamma$}

For the decay rate $\Gamma \sim \alpha_s(\Delta E) \, \Delta E^3\times[\text{multipole expansion}{\text{ + relativistic expansion} }]$, we consider uncertainties arising from:

\begin{itemize}
\item The transition energy $\Delta E\pm\delta\Delta E$, where $\delta\Delta E$ is the uncertainty in the binding energies discussed in Sec. \ref{sec:uncertainity_mass}.
\item  Higher orders in the strong coupling constant $\alpha_s$, $\delta \alpha_s=\frac{\alpha_s(\Delta E/2)-\alpha_s(2\Delta E)}{2}$, see Sec. \ref{sec:uncertainities_alpha}. This error includes the one due to the variation of $\alpha_s$ from $\delta\Delta E$.

\item Higher orders in the multipole expansion, {see Sec. \ref{sec:mul}.}

\item The use of eigenstates of the full (confining) potential in Eq. \eqref{cornell} rather than eigenstates of the singlet (perturbative) potential in  Eq. \eqref{pnrqcd}, see Sec. \ref{sec:Co}. This is estimated by calculating the average of the 
confining potential in the quarkonium state over the transition energy.

\item { Relativistic corrections, see Sec. \ref{sec:rel}. There are two contributions, one proportional to the energy tranfer ($\Delta E$) and one proportional to the average kinetic energy of the quark-antiquark pair in the initial and final states $\left<i|{\bf \partial}_r^2|f\right>$,} {we use the notation {${\cal V}\equiv   \frac{2}{m_Q} |\left<i|{\bf \partial}_r^2|f\right>|\simeq E_i-\left<i|V_h|i\right>+ E_f-\left<f|V_{\Sigma_g^+}|f\right> $,  $h=\Sigma_u^-\,,\Pi_u$, see} Eq. \eqref{relsc2}}.
\end{itemize}

 We add in quadrature the above errors and estimate the uncertainty  of the decay width as follows,\\

{\bf Spin Conserved:}

\begin{align}\label{eq:deltaGamma}
\frac{\delta \Gamma}{\Gamma} \sim \sqrt{ 
\left( 3 \, \frac{\delta \Delta E}{\Delta E} \right)^2 
+\left( \frac{\delta \alpha_s}{\alpha_s} \right)^2 
+ \left( \frac{|\langle i| r|f \rangle|^2 \Delta E^2}{120} \right)^2 
+ \left( 3\, \frac{\sigma {|\langle f| r|f \rangle|}+\mu}{\Delta E} \right)^2 +{ \left(\frac{\Delta E^2}{16 m_Q^2}\right)^2 + \left(\frac{\cal V}{8m_Q}\right)^2} }\,.
\end{align}

{\bf Spin Flip:}

\begin{align}\label{eq:deltaGamma2}
\frac{\delta \Gamma}{\Gamma} \sim \sqrt{ 
\left( 3 \, \frac{\delta \Delta E}{\Delta E} \right)^2 
+\left( \frac{\delta \alpha_s}{\alpha_s} \right)^2 
+ \left( \frac{|\langle i|r|f \rangle|^2 \Delta E^2}{24} \right)^2 
+ \left( 3\, \frac{\sigma {|\langle f|r|f \rangle|+\mu}}{\Delta E} \right)^2 +{\left(\frac{\Delta E^2}{16 m_Q^2}\right)^2 + \left(\frac{5\cal V}{24m_Q}\right)^2} }.
\end{align}

{We use 
\be \label{eq:irf_frf}
 |\langle i|r|f \rangle|=
 |\sum_L\int P^L_J(r)  r\ R_{L'}(r) \ dr |\quad,\quad |\langle f|r|f \rangle|=
 \int  r\ R_{L'}^2(r) \ dr \,.
\ee
Note that in contrast with Eq. \eqref{eq:<r>_cons}, Eq. \eqref{eq:irf_frf} does not include Clebsch-Gordan factors, see Sec. \ref{sec:mul}.}
 We  estimate $\left<i|V_h|i\right>$, $h=\Sigma_u^-\,,\Pi_u$ as follows, $\left<i|V_h|i\right>\sim V_h(r_\text{min})$,  the minimum of the average potential for the hybrid initial state, and $\left<f|V_{\Sigma_g^+}|f\right>\sim V_{\Sigma_g^+}(\left<f|r|f\right>)$ for the quarkonium final state. $V_h( r_\text{min}) \equiv \frac{1}{2}\text{Min}[V_{\Sigma^-_u}+V_{\Pi_u}]
\simeq 2m_Q+E_g^{q\bar{q}}+1.2$ GeV, where the numerical factor is estimated by analyzing the lattice data associated to the potentials used to compute hybrid and quarkonium states, see Fig. \ref{fig:potentials}.

\subsubsection{Uncertainty in the strong coupling constant} \label{sec:uncertainities_alpha} 

The strong coupling constant $\alpha_s(\Delta E)$ is evaluated at the scale of the energy difference between the hybrid and quarkonium states, we use the \texttt{AlphasExact} function from the \texttt{RunDec} package \cite{Chetyrkin:2000yt,Schmidt:2012az,Herren:2017osy} with 5 fixed flavor number and 4-loop  running, corresponding to evolution from $M_Z$ down to $\Delta E$. This setup ensures consistency with the flavor scheme and scale hierarchy adopted in Ref.~\cite{Brambilla:2022hhi}. To estimate the uncertainty on $\alpha_s(\Delta E)$, we vary the renormalization scale in the range $[\Delta E/2,\, 2\Delta E]$. This accounts for both the uncertainty in the energy gap $\Delta E$ and the effect of higher-order corrections in the running of $\alpha_s$. The resulting variation in $\alpha_s$ is then propagated into the decay width as a theoretical uncertainty.

\subsubsection{Higher orders in the multipole expansion}
\label{sec:mul}
For the spin-conserved case, higher order terms in the multipole expansion that we need to estimate the error can be found in \cite{Brambilla:2002nu}
\be
\mathcal{L}_{\rm multipole}^{sc}={\rm Tr} \left\{\!   
{\rm O}^\dagger {\bf r} \cdot g{\bf E}\,{\rm S} 
+{1 \over 24} 
{\rm O}^\dagger {\bf r}^i {\bf r}^j
  {\bf r}^k \, g {\bf 
    D}^i{\bf D}^j {\bf E}^k \,{\rm S} + \hbox{H.c.}\right\}
\ee
Recall that ${\bf D}$ and ${\bf E}$ depend only on the center of mass coordinate ${\bf R}$. The $\Delta L=1$ piece of ${\bf r}^i {\bf r}^j{\bf r}^k$ reads $r^2(\delta^{ij}{\bf r}^k +\delta^{jk}{\bf r}^i +\delta^{ki}{\bf r}^j)/5$. At leading order in $g$, ${\bf D}^i{\bf D}^j {\bf E}^k \simeq -\partial_i \partial_j \partial_0 A^k$. Since $A^k$ is transverse, the only remaining contraction is $-r^2 {\bf r}\,\boldsymbol{\partial}^2 \partial_0  {\bf A}/5, $ 
  The relative size of the higher order term is then,
  \be
  \frac{(\Delta E)^2|\langle f| r^3|i\rangle|}{120 |\langle f| r|i\rangle|}\simeq \frac{(\Delta E)^2|\langle f| r|i\rangle|^2}{120}
  \label{mulsc}
  \ee
For the spin-flip case, we have
\be
\mathcal{L}_{\rm multipole}^{sf}={c_F \over 2m} {\rm Tr} \left \{ {\rm O}^\dagger (\boldsymbol{\sigma}_1 - \boldsymbol{\sigma}_2) \left( g{\bf B} +\frac{{\bf r}^i{\bf r}^j}{8} {\bf D}^i{\bf D}^j g{\bf B}\right) \,{\rm S} 
+ \hbox{H.c.} \right \} \,.
\ee
Recall that ${\bf B}$ depends only on the center of mass coordinate ${\bf R}$. The $\Delta L=0$ piece of ${\bf r}^i {\bf r}^j$ reads $r^2 \delta^{ij}/3$, and at leading order in $g$ of ${\bf D}^i{\bf D}^j g{\bf B}\simeq \partial_i\partial_j \,g{\bf B}$. Hence the relevant piece of the correction reads $r^2 \boldsymbol{\partial}^2 g{\bf B}/24$, which leads to a higher order term of the following size,
\be
\frac{(\Delta E)^2|\langle f| r^2|i\rangle|}{24} \simeq \frac{(\Delta E)^2|\langle f| r|i\rangle|^2}{24}
\label{mulsf}
\ee

\subsubsection{Confining versus Coulomb potential}
\label{sec:Co}
The diagram in Fig. \ref{decay} produces an imaginary part to the hybrid potential, which is related to the total decay width to quarkonium plus anything else. This imaginary part {is calculated in weak coupling pNRQCD} and is proportional to \cite{Oncala:2017hop},
\be
\mathrm{Im}(V)\sim (E-h_s)^3 \quad,\quad h_s=\frac{{\bf p}^2}{m_Q}-\frac{C_f\als}{r}\,,
\ee
and $E$ is the energy of the hybrid. In the case of spin-conserved transitions this expression appears between two $r^i$, and in the case of spin-flip transitions between two $(\sigma^i_1-\sigma_2^i)/m_Q$. In order to estimate the contribution of a single quarkonium state to the decay width, we can project it into this state, say $|n\rangle$. However, our quarkonium states are {calculated in the (LO) BOEFT with a confining potential given in Eq. \eqref{VLambdaeta}, which corresponds to the strong coupling regime of pNRQCD. They are}{ approximate eigenstates of the Cornell potential $H_s=h_s+\sigma r+\mu$, $H_s |n\rangle=E_n |n\rangle$, with  $\mu=E_0-E_{J^{PC}}=-0.07/r_0=-28$ MeV and   $\sigma =0.215\, \mathrm{GeV}^2$. $H_s$ is a very good approximation to the confining potential in Eq. \eqref{VLambdaeta},
see
Fig. \ref{fig:potentials}. 
For the lowest lying states the confining term can be treated as a perturbation,
\be
\langle n|(E-h_s)^3|n\rangle=\langle n|(E-H_s+\sigma r)^3|n\rangle\simeq (E-E_n)^3 +3(E-E_n)^2\left[ \sigma\langle n| r  |n\rangle+\mu\right]
\ee
Hence the relative error made by replacing the Coulomb potential by the Cornell potential is $3\left[ \sigma \langle f|r|f\rangle+\mu\right]/\Delta E$, where we have identified $|n\rangle$ with the final quarkonium state $|f\rangle$. {This error accounts for the difference between weak and strong coupling regimes.}

\subsubsection{Relativistic corrections}
\label{sec:rel}

The relativistic corrections are better discussed at the NRQCD level. For the quark piece of the spin-conserved transitions we have,
\be
{\cal L}_{\rm relativistic}^{sc}= \psi^\dagger \Biggl\{ 
{{\bf D}^2\over 2 m_Q}-{{\bf D}^4\over 8 m_Q^3} 
\Biggr\} \psi\simeq \psi^\dagger \Biggl\{ 
{\left\{\boldsymbol{\partial}\,, ig\boldsymbol{A}\right\}\over 2 m_Q}-{\left\{\left\{\boldsymbol{\partial}\,, ig\boldsymbol{A}\right\}\,,\boldsymbol{\partial}^2\right\}\over 8 m_Q^3} 
\Biggr\}\psi\,,
\ee
where we only keep the terms contributing to the leading order in the multipole expansion. Upon projection to the quark-antiquark sector \cite{Pineda:1997bj}, we have (time dependence is implicit),
\bea
{\cal L}_{\rm relativistic}^{sc} &=& 
\psi^\dagger(\boldsymbol{x}_1,\boldsymbol{x}_2)\left( {\left\{\boldsymbol{\partial}_1\,, ig\boldsymbol{A}(\boldsymbol{x}_1)\right\}\over 2 m_Q}-{\left\{\left\{\boldsymbol{\partial}_1\,, ig\boldsymbol{A}(\boldsymbol{x}_1)\right\}\,,\boldsymbol{\partial}_1^2\right\}\over 8 m_Q^3} \right.\\ 
&& \left. - {\left\{\boldsymbol{\partial}_2\,, ig\boldsymbol{A}(\boldsymbol{x}_2)\right\}\over 2 m_Q}+{\left\{\left\{\boldsymbol{\partial}_2\,, ig\boldsymbol{A}(\boldsymbol{x}_2)\right\}\,,\boldsymbol{\partial}^2_2\right\}\over 8 m_Q^3}\right)\psi(\boldsymbol{x}_1,\boldsymbol{x}_2)\,.\nn
\eea
At leading order in the multipole expansion $\boldsymbol{A}(\boldsymbol{x}_1)\simeq \boldsymbol{A}(\boldsymbol{x}_2)  \simeq \boldsymbol{A}(\boldsymbol{R})$. In terms of the center of mass $\boldsymbol{R}$ and relative coordinate $\boldsymbol{r}$, $\boldsymbol{\partial}_1=\boldsymbol{\partial}_R/2+\boldsymbol{\partial}_r$, $\boldsymbol{\partial}_2=\boldsymbol{\partial}_R/2-\boldsymbol{\partial}_r$

\be
{\cal L}_{\rm relativistic}^{sc}= 
\psi^\dagger(\boldsymbol{R},\boldsymbol{r})\left( {2\boldsymbol{\partial}_r\, ig\boldsymbol{A}(\boldsymbol{R})\over m_Q}-{\boldsymbol{\partial}_R^2\boldsymbol{\partial}_r\, ig\boldsymbol{A}(\boldsymbol{R})\over 8 m_Q^3} -{4\boldsymbol{\partial}_r\, ig\boldsymbol{A}(\boldsymbol{R})\boldsymbol{\partial}^2_r\over 8 m_Q^3}\right)\psi(\boldsymbol{R},\boldsymbol{r})\,,
\label{rel}
\ee
where we have used that the gluon field is transverse and assumed that eventually the initial state will be at rest. The middle term has the same dependence in the relative coordinate as the leading one. Hence it leads to a correction of relative size,
\be
\frac{(\Delta E)^2}{16 m_Q^2}
\label{relsc1}
\ee
The last term in Eq. \eqref{rel} is estimated as follows
\be
|\langle f| \boldsymbol{\partial}_r \boldsymbol{\partial}_r^2 |i\rangle| \simeq |\frac{\langle f| \boldsymbol{\partial}_r |i\rangle|| \langle i|\boldsymbol{\partial}_r^2 |i\rangle|+|\langle f|  \boldsymbol{\partial}_r^2 |f\rangle|| \langle f|\boldsymbol{\partial}_r|i\rangle|}{2}
\equiv \frac{|\langle f|\boldsymbol{\partial}_r|i\rangle| m_Q \mathcal{V}}{2}
\label{relsc2}
\ee
The relevant terms for calculating the relativistic corrections to the spin-flip transitions can be found in \cite{Manohar:1997qy}. For the quark sector they read,
\be
{\cal L}_{\rm relativistic}^{sf}= \psi^\dagger \Biggl\{
\, g {{\bf \boldsymbol{\sigma} \cdot B} \over 2 m_Q} 
 + 
\, g { \left\{ {\bf D^2,\boldsymbol{\sigma}
\cdot B }\right\}\over 8 m_Q^3} - \, g { {\bf D}^i\, {\bf \boldsymbol{\sigma}
\cdot B} \, {\bf D}^i  \over 4 m_Q^3} 
+ \, g { {\bf \boldsymbol{\sigma} \cdot D\, B \cdot D + D 
\cdot B\, \boldsymbol{\sigma} \cdot D}\over 8 m_Q^3} \Biggl\}\psi
\label{manohar}
\ee
Following the same procedure as for the spin-conserved ones, we obtain for the quark-antiquark sector,

\be
{\cal L}_{\rm relativistic}^{sf}= 
\psi^\dagger(\boldsymbol{R},\boldsymbol{r})\left( \frac{(\boldsymbol{\sigma}_1 - \boldsymbol{\sigma}_2)  g{\bf B}}{2m_Q}+\frac{(\boldsymbol{\sigma}_1 - \boldsymbol{\sigma}_2) \boldsymbol{\partial}_R^2 g{\bf B}}{32m_Q^3}+\frac{5\,(\boldsymbol{\sigma}_1 - \boldsymbol{\sigma}_2) \boldsymbol{\partial}_r^2 g{\bf B}}{24m_Q^3}
\right)\psi(\boldsymbol{R},\boldsymbol{r})\,.
\ee
We have focused on corrections to the leading term. In particular, we have projected onto the $\Delta L=0$ sector the contributions arising from the last term in Eq. \eqref{manohar}. The term in the middle leads to a correction,
\be
\frac{(\Delta E)^2}{16m_Q^2}
\label{relsf1}
\ee 
The size of the last term is estimated as follows,
\be
|\langle f| \boldsymbol{\partial}_r^2 |i\rangle| \simeq \frac{ |\langle i|\boldsymbol{\partial}_r^2 |i\rangle|+|\langle f|  \boldsymbol{\partial}_r^2 |f\rangle| }{2}=\frac{m_Q\mathcal{V}}{2}
\label{relsf2}
\ee 

\section{Potentials\label{sec:stat_pot}}

The spectrum of quarkonium states can be obtained by solving a Schr\"odinger equation with an appropriate potential. This potential generally incorporates non-perturbative QCD dynamics and thus cannot be derived using perturbation theory alone. Non-perturbative methods, such as lattice QCD, have been employed to compute the static potential in the intermediate regime~\cite{PhysRevD.62.054503,Juge:2002br
}. In the short-distance region ($r \ll 1/\Lambda_{\text{QCD}}$), perturbative QCD is reliable and can be applied successfully~\cite{PINEDA2012735}. At long distances ($r \gg 1/\Lambda_{\text{QCD}}$), the gluonic field between the heavy quarks forms a flux tube, whose dynamics can be described using effective string theory~\cite{PhysRevD.79.114002,Hwang:2018rju}.

To span the entire range of quark-antiquark separations, we use piecewise-defined potentials that interpolate between the short- and long- distance behaviors. These potentials are fitted to recent lattice QCD results for the static energies~\cite{Alasiri:2024nue}.
\begin{align}
\tilde{V}_{\Lambda_\eta^\epsilon}(r)  &=
 \frac{\kappa}{r}+ E_{J^{PC}}  +  A_{\Lambda_\eta^\epsilon} \,r^2  +  B_{\Lambda_\eta^\epsilon} \,r^4 + C_{\Lambda_\eta^\epsilon} \,r^6
\qquad  r < r_{\Lambda_\eta^\epsilon},\nn
\\
 &=  V_N(r)  +E_0+  \frac{ D_{\Lambda_\eta^\epsilon}}{r^4}  
 \qquad\qquad\qquad\qquad\, ~~~ \ \   r > r_{\Lambda_\eta^\epsilon},
\label{VLambdaeta}%
\end{align}
 corresponding to $\Sigma_g^+$, $\Pi_u$ and $\Sigma_u^-$. The potential of the string is
$
V_N(r) = \sqrt{\sigma^2\, r^2 + 2 \pi \big(N - \tfrac{1}{12}\big) \sigma},
\label{V-N}
$
$\sigma$ is the string tension, and $N$ is the quantum number for transverse vibrational excitations of the string. $E_{J^{PC}}$ is the energy of the gluelump with quantum numbers ${J^{PC}}$. The potential parameters are summarized in Table \ref{tab:SigmagpPiu-params}. We find that the energy shift
\begin{equation}
    V_{\Lambda_\eta^\epsilon}(r) = \tilde{V}_{\Lambda_\eta^\epsilon}(r) + 2m_Q + E_g^{Q\bar{Q}}
\end{equation}
that yields the best agreement between the predicted and experimental spectra for both charmonium and bottomonium. Using $m_c = 1.47~\text{GeV}$ for the charm quark mass and $m_b = 4.88~\text{GeV}$ for the bottom, we extract the gluonic energy shifts as $E_g^{c\bar{c}} = -428~\text{MeV}$ for charmonium and $E_g^{b\bar{b}} = -423~\text{MeV}$ for bottomonium.

\begin{table}[h]
\centering
\resizebox{0.8\textwidth}{!}{%
\begin{tabular}{|c c c|c c c|c c c c c c|}
\hline
\rowcolor{cyan!20}
$J^{PC}$ & $\Lambda_\eta^\epsilon$ & $N$ &
$r_0^2 \sigma$ & $r_0 E_0$ & $\kappa$ &
$r_0 E_{J^{PC}}$ & $r_0^3 A$ & $r_0^5 B$ & $r_0^7 C$ & $D/r_0^3$ & $r_{\Lambda_\eta^\epsilon}/r_0$ \\
\hline
$0^{++}$ & $\Sigma_g^+$ & 0 & 1.384 & -0.057 & -0.240 & 0.013 & 3.20 & -4.69 & 3.49493 & 0.02874 & 0.713 \\
$1^{+-}$ & $\Pi_u$       & 1 & 1.384 & -0.057 &  0.037 & 2.984 & -0.010 &  0.209 & -0.05314 & 0.12717 & 1.298 \\
$1^{+-}$ & $\Sigma_u^-$ & 3 & 1.384 & -0.057 &  0.037 & 2.984 & 0.683 & 0.009 & -0.03246 & -16.96753 & 2.216 \\
\hline
\end{tabular}
}
\caption{\small
Dimensionless parameters for the $\Sigma_g^+$, $\Pi_u$, and $\Sigma_u^-$ potentials in pure $SU(3)$ gauge theory. Where the Sommer scale $r_0 = 0.5~\text{fm}$ and hence the string tension $\sigma =0.215\, \mathrm{GeV}^2$ is used throughout. Since the cutoff in the potentials are established at different points, \(V_{\Sigma_u^-}\) shall be defined from its difference with \(V_{\Pi_u}\) i.e. \(V_{\Sigma_u^-} = V_{\Pi_u} + \Delta V_{\Sigma_u^-}\), as determined in~\cite{Alasiri:2024nue}. 
\label{tab:SigmagpPiu-params}}
\end{table}

\begin{figure}[H]
    \centering
    \includegraphics[width=0.425\linewidth]{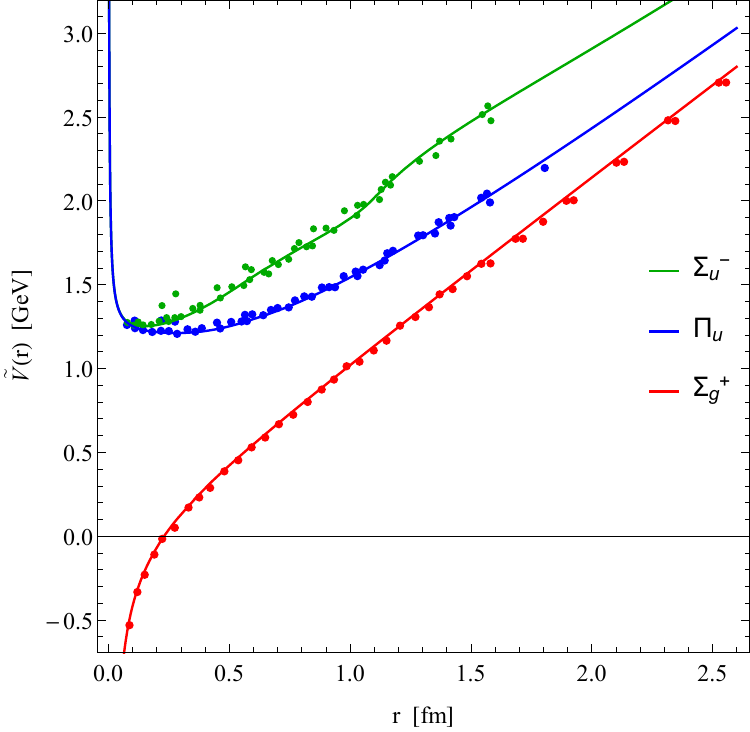}
    \includegraphics[width=0.43\linewidth]{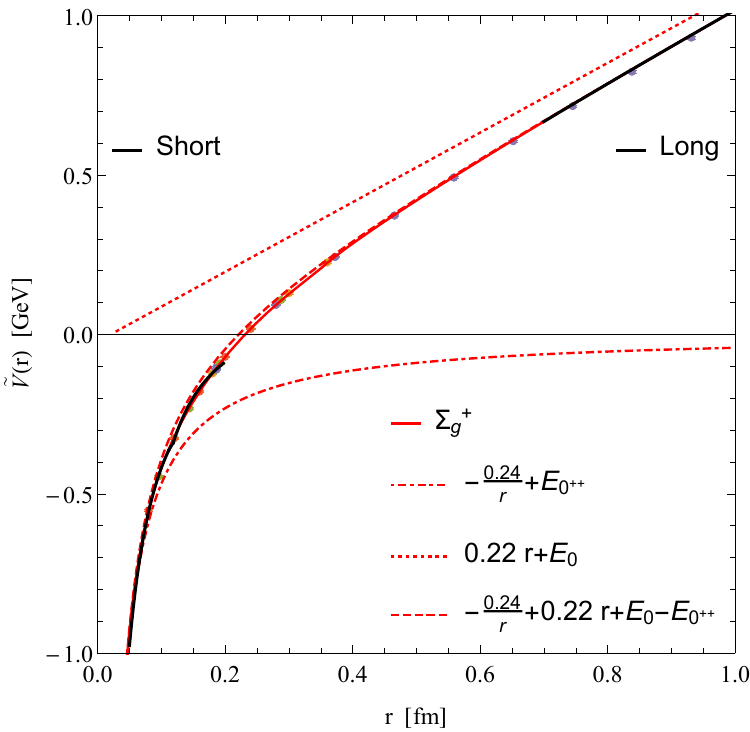}
    \caption{\footnotesize\label{fig:potentials} { Left panel:} the {\it blue} and {\it green lines} show the fit of the hybrid potentials $\Pi_u$ and $\Sigma_u$, respectively. The {\it solid red line} shows the fit of the static potential $\Sigma_g^+$ in Eq. \eqref{VLambdaeta} proposed in ref. \cite{Alasiri:2024nue} on top of the lattice data ({\it dots}) from  ref. \cite{Capitani:2018rox,Schlosser:2021wnr}. The energy gap of $\sim1.2$ GeV 
    {corresponds to the energy difference between} the minimum of the hybrid potentials {and the value of the $\Sigma_g^+$ potential at the distance where the minimum occurs}. {Right panel: the {\it red dash-dotted lines} show the Coulomb potential and  the {\it red dotted lines} the confining potential} in order to illustrate the energy gap between long- and short- distance regimens, $\mu=E_0-E_{0^{++}}=-28$ MeV. We also show in black the expected long (EST) and short (perturbative) distance behavior. The {\it red dashed line} shows the corresponding Cornell approximation that we use in section \ref{sec:Co} to estimate the error.} 
\end{figure}
\clearpage
\section{Scalar, Vector and Tensor Spherical Harmonic properties}
\label{Ap.VSH}

Quarkonium (spin 0) wave-functions can be expressed in terms of scalar spherical harmonics (SH),
\begin{equation}
\label{S}
S\equiv  
\sum_{L=0}^{\infty}\sum_{M=-L}^{L}  
\frac{R_L(r)}{r}
{Y}^{L}_{M}   
\end{equation}
while the spin 1 quarkonium wave-function has a vector form and shall be expressed in terms of  vector spherical harmonics (VSH)
\begin{equation}
\label{S1}
S^{i}= \sum_{J=0}^{\infty}\sum_{M=-J}^{J} { Y^i}^{L}_{JM} \frac{R_{L}(r)}{r}
\end{equation}
where $L$ can take the values of $L=|J-1|,J,J+1$ for $J>0$, and only $L=J+1$ for $J=0$. Hybrid (spin 0) wave-function has also a vector form and admit the follow decomposition \cite{Oncala:2017hop}:
\begin{equation}
\label{H}
H^i=
\sum_{J=0}^{\infty}\sum_{M=-J}^{J}  
\frac{1}{r}
\left[
 P_J^{+} (r) { Y^i}_{JM}^{L=J+1}   
+P_J^{0}(r) {Y^i}_{JM}^{L=J}  
+P_J^{-}(r) {Y^i}_{JM}^{L=|J-1|}
\right]
\end{equation}
 where $P_J^{L}(r)$ is the radial $L=J+1$ and transversal $L=J,|J-1|$ components of the vector field. We recall that for $J = 0$, only the radial component $L = J + 1 = 1$ contributes. Vector spherical harmonics ${ Y^i}_{JM}^L$ can be expanded in terms of scalar spherical harmonics ${Y}^{L}_{M}$ using Clebsch-Gordan coefficients. 
\begin{equation}
    {Y^i}_{JM}^{L}=\sum_{\mu} C(L,1,J;M-\mu,\mu)Y^{L}_ {M-\mu}\chi^i_\mu
\label{expansion_VSH}
\end{equation}
 The index $i$ is the one from the VSH. $\mu$ sums for $\{-1,0,1\}$, while $C$ are Clebsch-Gordan coefficients and $ \chi^i_\mu$ is the unitary director vector base of VSH, defined as:
\begin{equation}
\label{e}
{\chi}_{+1}=-\frac{\widehat{x}+i\widehat{y}}{\sqrt{2}}, 
\hspace{1cm} 
{\chi}_0=\widehat{z}, 
\hspace{1cm} 
{\chi}_{-1}=\frac{\widehat{x}-i\widehat{y}}{\sqrt{2}}.
\end{equation}
Vector spherical harmonics obey orthogonality
\begin{equation}
\label{orto}
\int_{\Omega}  {{ Y^i}^*}_{J'M'}^{L'}  { Y^j}_{JM}^L 
=\delta^{L}_{L'} \ \delta^{J}_{J'} \ \delta^{M}_{M'} \ \delta_i^j
\end{equation}
The radial projection on the spherical harmonics is:
\begin{equation}
    \widehat{r}^i \cdot Y^{L}_{M}\equiv -\sum_{l}C(L,1,l;0,0){ Y^i}^{l}_{LM}=-\sqrt{\frac{L+1}{2L+1}}{Y^i}^{L+1}_{LM}+\sqrt{\frac{L}{2L+1}}{ Y^i}^{|L-1|}_{LM}
    \label{rY}
\end{equation}
While the radial projection on the vector spherical harmonics are:
\begin{align}
\widehat{r}^i \ { Y^j}_{JM}^{J}=0
\qquad\qquad \qquad
\widehat{r}^i \ { Y^j}_{JM}^{J+1}=-\sqrt{\frac{J+1}{2J+1}} \ Y^J_M
\qquad \qquad \qquad
\widehat{r}^i \  { Y^j}_{JM}^{|J-1|}=\sqrt{\frac{J}{2J+1}}Y^J_M
\end{align}
We can express the wave-function for the hybrid state with spin 1 in terms of tensor spherical harmonics:
\begin{equation}
\label{TensorH}
H^{ji}=\sum_{\j=0}^{\infty} \sum_{\m=-\j}^{\j} \frac{P^{LJ}_{\j}(r)}{r}  \ Y^{ji\, LJ}_{\j\m} 
\end{equation}
For a given $\j$, we get $J=\j-1,\j,\j+1$ and $L=J-1,J,J+1$ components. Notice that for $\j=0$ only $J=\j+1=1$ component exists, and for $J=0$ only $L=J+1$ component exist. Tensor Spherical Harmonics are defined following the notation in \cite{pg} as:
\begin{equation}
\label{TSH}
Y_{\cal JM}^{ijLJ} =\sum_{\nu} C(J,1,{\cal J}|{\cal M}-\nu,\nu)Y_{J {\cal M}-\nu}^{i L} {\chi}_\nu^j
\end{equation}
and so using Eq. \eqref{expansion_VSH}:
\begin{equation}
\label{TSH2}
Y_{\cal JM}^{ijLJ} =\sum_{\nu\mu} C(J,1,{\cal J}|{\cal M}-\nu,\nu)
C(L,1,J|{\cal M}-\nu-\mu,\mu)Y^L_{\m-\nu-\mu} 
{\chi}_\mu^i{\chi}_\nu^j
\end{equation}

\section{Spin-flip matrix element \label{app:spin_fliped_element}}

Being $\vec{s}_{1}=\vec{\sigma}/2$ and $\vec{s}_{2}=-\vec{\sigma}^T/2$ the Pauli matrices, and using the notation ${a^T}^{\ j}_{j'}=a^{j}_{\ j'}$, the total spin operator can be written as
\begin{equation}
  \vec{s}_{1}-\vec{s}_{2}=
  \left(\frac{\vec{\sigma}}{2}\right)^{i'}_{\ i}\delta^{\ j}_{j'}
  -
  \delta^{i'}_{\ i}\left(-\frac{\vec{\sigma}^T}{2}\right)^{\ j}_{j'}
  =
  \frac{1}{2}\left(
  \vec{\sigma}^{i'}_{\ i}\delta^{j}_{\ j'}
  +
  \delta^{i'}_{\ i}\vec{\sigma}^{j}_{\ j'}\right).
\end{equation}
Considering
\begin{align}
    H_J&=\frac{1}{\sqrt{2}}\left(H^n+H^{nl}\sigma^l\right)^i_{\ j},\\
    S_{L'}^*&=\frac{1}{\sqrt{2}}\left(S^*+{S^{l'}}^*\sigma^{l'}\right)^{j'}_{\ i'}.
\end{align}
The matrix element is
\begin{align}
    \label{matrix_element_spin_flip}
    \left<S_{L'}|\vec{s}_1-\vec{s}_2|H_J\right>&=
    \frac{1}{4}
    \left(S^*\delta^{j'}_{\ i'}+{S^{l'}}^*{\sigma^{l'}}^{j'}_{\ i'}\right)
    \left(
  \vec{\sigma_1}^{i'}_{\ i}\delta^{j}_{\ j'}
  +
  \delta^{i'}_{\ i}\vec{\sigma_2}^{j}_{\ j'}\right)
  \left(H^n\delta^i_{\ j}+H^{nl}{\sigma^l}^i_{\ j}\right)=\\
    &{=\frac{1}{4}
    \left(S^*\delta^{j'}_{\ i'}+{S^{l'}}^*{\sigma^{l'}}^{j'}_{\ i'}\right)
    \left(H^n(\vec{\sigma_1}+\vec{\sigma_2})^{i'}_{\ j'}+
    H^{nl}(\vec{\sigma_1} \sigma^l+\sigma^l\vec{\sigma_2})^{i'}_{\ j'}\right)=}\nonumber\\
    &=\frac{1}{4}\left(S^*H^n(\vec{\sigma_1}+\vec{\sigma_2})^{j'}_{\ j'}+S^*H^{nl}(\vec{\sigma_1} \sigma^l+\sigma^l\vec{\sigma_2})^{j'}_{\ j'}\right)+\nonumber\\
    &+\frac{1}{4}\left(S^{l'*}H^n(\sigma^{l'}(\vec{\sigma_1}+\vec{\sigma_2})^{j'}_{\ j})+S^{l'*}H^{nl}(\sigma^{l'}(\vec{\sigma_1} \sigma^l+\sigma^l\vec{\sigma_2}))^{j'}_{\ j'}\right)=\nonumber\\
    &={S^*\vec{H}^{n}+\vec{S}^{*}H^n \,,}\nonumber
\end{align}
{where $(\vec{H}^n)^k=H^{nk}$ and $(\vec{S})^k=S^k$.} From Eq.  \eqref{matrix_element_spin_flip} we see transitions from hybrid spin 0  to quarkonium spin 1 ($S^{k*}H^n$) and from hybrid spin 1 to quarkonium spin 0 ($S^{*}H^{nk}$).
\newpage

\bibliographystyle{ieeetr}
\bibliography{bibliography.bib}

\end{document}